\documentclass[twocolumn]{revtex4}
\usepackage[dvips]{graphicx}
\usepackage{float, color,array,amsmath}

\begin{document}

\title{Effects of magnetic anisotropy on spin and thermal transports in classical antiferromagnets on the square lattice}

\author{Kazushi Aoyama and Hikaru Kawamura}

\date{\today}

\affiliation{Department of Earth and Space Science, Graduate School of Science, Osaka University, Osaka 560-0043, Japan
}

\begin{abstract}
Transport properties of the classical antiferromagnetic XXZ model on the square lattice have been theoretically investigated, putting emphasis on how the occurrence of a phase transition is reflected in spin and thermal transports. As is well known, the anisotropy of the exchange interaction $\Delta\equiv J_z/J_x$ plays a role to control the universality class of the transition of the model, i.e., either a second-order transition at $T_N$ into a magnetically ordered state or the Kosterlitz-Thouless (KT) transition at $T_{KT}$, which respectively occur for the Ising-type ($\Delta >1$) and $XY$-type ($\Delta <1$) anisotropies, while for the isotropic Heisenberg case of $\Delta=1$, a phase transition does not occur at any finite temperature. 
It is found by means of the hybrid Monte-Carlo and spin-dynamics simulations that the spin current probes the difference in the ordering properties, while the thermal current does not. For the $XY$-type anisotropy, the longitudinal spin-current conductivity $\sigma^s_{xx}$ ($=\sigma^s_{yy}$) exhibits a divergence at $T_{KT}$ of the exponential form, $\sigma^s_{xx} \propto \exp\big[ B/\sqrt{T/T_{KT}-1 }\, \big]$ with $B={\cal O}(1)$, while for the Ising-type anisotropy, the temperature dependence of $\sigma^s_{xx}$ is almost monotonic without showing a clear anomaly at $T_{N}$ and such a monotonic behavior is also the case in the Heisenberg-type spin system. The significant enhancement of $\sigma^s_{xx}$ at $T_{KT}$ is found to be due to the exponential rapid growth of the spin-current-relaxation time toward $T_{KT}$, which can be understood as a manifestation of the topological nature of a vortex whose lifetime is expected to get longer toward $T_{KT}$. Possible experimental platforms for the spin-transport phenomena associated with the KT topological transition are discussed.      
\end{abstract}

\maketitle
\section{Introduction}
Transport phenomena in magnetic systems reflect dynamical properties of interacting spins, such as magnetic excitations and fluctuations. Of recent particular interest is spin transport which is becoming available as a probe to study magnetic properties thanks to the development of experimental methods in the context of spintronics \cite{Spincurrent-mag_Frangou_16, Spincurrent-mag_Qiu_16, Spincurrent-mag_Wang_17, Spincurrent-mag_Frangou_17, Spincurrent-mag_Gladii_18, Spincurrent-mag_Ou_18}. This demands to explore the fundamental physics underlying the association between the spin transport and magnetic phase transitions.
In this paper, we theoretically investigate transport properties of two-dimensional antiferromagnetic insulators, putting emphasis on the effects of magnetic anisotropy which plays a role of controlling the universality class of the system.

A minimal model of two-dimensional antiferromagnets with magnetic anisotropy would be the classical nearest-neighbor (NN) antiferromagnetic XXZ model on the square lattice. The spin Hamiltonian of the system is given by 
\begin{equation}\label{eq:Hamiltonian}
{\cal H} = - J\sum_{\langle i,j \rangle} \Big( S^x_i S^x_j + S^y_i S^y_j + \Delta S^z_i S^z_j  \Big), 
\end{equation}
where $S^\alpha_i$ is $\alpha$-component of a classical spin ${\bf S}_i$ at a lattice site $i$, $\langle i,j\rangle$ denotes the summation over all the NN pairs, $J< 0$ is an antiferromagnetic exchange interaction, and $\Delta > 0$ is a dimensionless parameter characterizing the magnetic anisotropy. The ground state of this system is the conventional two-sublattice antiferromagnetic order, whereas the finite-temperature properties depend on the magnetic anisotropy $\Delta$. In the isotropic case of $\Delta=1$, Eq. (\ref{eq:Hamiltonian}) is nothing but the isotropic Heisenberg model, so that a phase transition does not occur at any finite temperature. In the anisotropic case of $\Delta > 1$ ($\Delta < 1$), the system belongs to the Ising ($XY$) universality class and exhibits a magnetic (Kosterlitz-Thouless topological \cite{KT_KT_73}) phase transition at a finite temperature $T_N$ ($T_{KT}$). The purpose of this work is to clarify how the difference in the ordering properties among the three cases, $\Delta >1$, $\Delta<1$, and $\Delta = 1$, is reflected in the transport properties. Our main focus is on whether a signature of a phase transition shows up in the spin and thermal transports or not. 

In the ordered phase at lower temperatures, spin and thermal currents should be carried by spin waves or magnons. With increasing temperature, thermally-activated nontrivial excitations and fluctuations would come into play. In particular, in the case of the $XY$-type anisotropy ($\Delta < 1$), free vortices dissociated at higher temperature above $T_{KT}$ may strongly affect the current relaxation, because the topological object of the vortex is generally robust against weak perturbations, resulting in a relatively long lifetime compared with the damping of the spin-wave mode \cite{SpnDyn-XY_Huber_82,SpnDyn-XY_Mertens_87,SpnDyn-XY_Gouvea_89,SpnDyn-XY_Evertz_96}. As we will demonstrate below, this is actually the case for the spin-current relaxation. In this paper, we will investigate temperature dependences of the conductivities of the spin and thermal currents in the Ising-type ($\Delta > 1$), $XY$-type ($\Delta < 1$), and Heisenberg-type ($\Delta = 1$) spin systems by means of the hybrid Monte-Carlo (MC) and spin-dynamics simulations.

Our result is summarized in Fig. \ref{fig:overview}. The longitudinal thermal conductivity $\kappa_{xx}$, which is the response to the temperature gradient $\nabla T$, is insensitive to the difference in the ordering properties. $\kappa_{xx}$ increases toward $T=0$ as a power function of temperature $T$ in all the three cases of $\Delta >1$, $\Delta <1$, and $\Delta =1$, without showing a clear anomaly at $T_N$ and $T_{KT}$. In contrast, the longitudinal spin-current conductivity $\sigma^s_{xx}$, which is the response to the magnetic-field gradient $\nabla H$, exhibits temperature dependences characteristic of the three different universality classes. For the $XY$-type anisotropy, $\sigma^s_{xx}$ exhibits a divergent sharp peak at $T_{KT}$, while for the Ising-type anisotropy, the temperature dependence of $\sigma^s_{xx}$ is monotonic without showing a clear anomaly at $T_{N}$. In the Heisenberg-type isotropic case, $\sigma^s_{xx}$ shows an exponential increase toward $T=0$. 
The significant enhancement of $\sigma^s_{xx}$ at $T_{KT}$ is due to the spin-current relaxation getting slower toward $T_{KT}$, which can be understood as a manifestation of the topological nature of the vortex whose lifetime is expected to get longer toward $T_{KT}$.

This paper is organized as follows: In Sec. II, the theoretical framework for transport phenomena in magnetic insulators will be given. We derive the expressions for the conductivities of the spin and thermal currents, and explain the details of our simulations. In Sec. III, low-temperature transport properties will be discussed based on analytical calculations within the linear spin-wave theory. Numerical results on the thermal and spin transports will be shown in Secs. IV and V, respectively. We end this paper with summary and discussions in Sec. VI.

\section{Theoretical framework for transport phenomena in magnets}
In this section, starting from the introduction to the equation of motion for the spin dynamics, we first derive the spin and thermal currents by using this spin-dynamics equation, and then, formulate the spin-current conductivity $\sigma^s_{\mu\nu}$ and the thermal conductivity $\kappa_{\mu\nu}$ within the linear response theory. Subsequently, we will explain numerical methods to integrate the equation of motion, taking account of temperature effects. 
\subsection{Spin dynamics}
The spin dynamics, i.e., the time evolution of the spins for the Hamiltonian (\ref{eq:Hamiltonian}), is determined by the following semiclassical equation of motion: 
\begin{eqnarray}\label{eq:Bloch}
\frac{d {\bf S}_i}{dt} &=&  {\bf S}_i \times {\bf H}_i^{\rm eff}, \nonumber\\
{\bf H}_i^{\rm eff} &=& J\sum_{j \in N(i)} \big( S^x_j, S^y_j, \Delta S^z_j  \big), 
\end{eqnarray}
where $N(i)$ denotes all the NN sites of $i$. Since Eq. (\ref{eq:Bloch}) is a classical analogue of the Bloch equation, namely, the Heisenberg equation for the spin operator, all the static and dynamical magnetic properties purely intrinsic to the Hamiltonian (\ref{eq:Hamiltonian}) should be described by the combined use of Eqs. (\ref{eq:Hamiltonian}) and (\ref{eq:Bloch}). A familiar alternative way to examine the spin dynamics is solving the Landau-Lifshitz-Gilbert (LLG) equation \cite{LLG_Landau_35} which includes a damping term originally introduced phenomenologically. In this work, we use Eq. (\ref{eq:Bloch}) instead of the LLG equation for the following two reasons: (i) In the LLG equation, the damping, which is characterized by a dimensionless parameter $\alpha$, may be either intrinsic or extrinsic to the spin Hamiltonian. Equation (\ref{eq:Bloch}), on the other hand, corresponds to the LLG equation without the phenomenological damping term, so that any relaxation described by Eq. (\ref{eq:Bloch}) has its origin in the Hamiltonian (\ref{eq:Hamiltonian}). As our focus in the present paper is on fundamental aspects intrinsic to the spin Hamiltonian (\ref{eq:Hamiltonian}), we use Eq. (\ref{eq:Bloch}); (ii) As we will see in the following subsection, the conventional forms of the spin and thermal currents \cite{SpinDyn_Huber_74, SpinDyn_Jencic_prb_15, MHall_Mook_prb_16, MHall_Mook_prb_17, Thermal_Huber_ptp_68, SpinDyn_Zotos_prb_05, SpinDyn_Sentef_07, SpinDyn_Pires_09, SpinDyn_Chen_13} are derived from the Heisenberg equation or its classical analogue without the damping term, so that it is self-consistent to use Eq. (\ref{eq:Bloch}) rather than the LLG equation with the additional damping term. 

\subsection{Conductivities of spin and thermal currents}
In this subsection, we will derive the spin current ${\bf J}_s^z$ and the thermal current ${\bf J}_{th}$, and then, introduce the spin-current conductivity $\sigma^s_{\mu\nu}$ and the thermal conductivity $\kappa_{\mu\nu}$. We shall start from the general discussion on a current in the continuum limit. Suppose that a conserved physical quantity ${\cal O}=\int d{\bf r} \,  {\cal O}({\bf r},t)$ should satisfy the continuity equation $\frac{\partial}{\partial t}{\cal O}({\bf r},t)+\nabla \cdot {\bf j}_{\cal O}({\bf r}, t)=0$ with associated local current density ${\bf j}_{\cal O}({\bf r}, t)$. By multiplying the both side of the equation by ${\bf r}$ and integrating over the whole ${\bf r}$ region, we obtain
\begin{equation}
\int d{\bf r} \, {\bf r} \frac{\partial}{\partial t}{\cal O}({\bf r},t) = -\int d{\bf r} \, {\bf r} \,\nabla \cdot {\bf j}_{\cal O}({\bf r}, t) = \int d{\bf r} \, {\bf j}_{\cal O}({\bf r}, t).
\end{equation} 
Thus, the net current ${\bf J}_{\cal O}(t)$ is given by \cite{book_Mahan}
\begin{equation}
 {\bf J}_{\cal O}(t)=\int d{\bf r} \, {\bf j}_{\cal O}({\bf r}, t) = \int d{\bf r} \, {\bf r} \frac{\partial}{\partial t}{\cal O}({\bf r},t).
\end{equation}
In the present XXZ model given by Eq. (\ref{eq:Hamiltonian}), the conserved quantities are the $z$ component of the magnetization $M^z=\sum_i S^z_i$ and the total energy ${\cal H} = \sum_i {\cal H}_i$ with ${\cal H}_i=\frac{-J}{2}\sum_{j \in N(i)}\big( S^x_i S^x_j + S^y_i S^y_j + \Delta S^z_i S^z_j  \big)$, so that the associated currents, namely, the spin and thermal currents (${\bf J}^z_s$ and ${\bf J}_{th}$) are given by 
\begin{eqnarray}\label{eq:j_sc}
{\bf J}^z_s(t) &=& \sum_i {\bf r}_i \frac{d S^z_i}{d t} \nonumber\\
&=& -J\sum_i {\bf r}_i \sum_{j \in N(i)} \big( {\bf S}_j \times {\bf S}_i \big)^z \nonumber\\
&=& J\sum_{\langle i,j \rangle} \big({\bf r}_i-{\bf r}_j \big) \big( {\bf S}_i \times {\bf S}_j \big)^z, 
\end{eqnarray}
\begin{eqnarray}\label{eq:j_th}
{\bf J}_{th}(t) &=& \sum_i {\bf r}_i \frac{-J}{2}\sum_{j \in N(i)}\frac{d}{dt}\big( S^x_iS^x_j+S^y_i S^y_j+\Delta S^z_i S^z_j \big) \nonumber\\
 &=& \frac{J^2}{2}\sum_i {\bf r}_i \sum_{j \in N(i)}\Big(\sum_{k \in N(i)}\Big\{({\bf S}_j\times{\bf S}_k)^zS^z_i \nonumber\\
 && + \Delta \Big[({\bf S}_i\times{\bf S}_j)^zS_k^z + ({\bf S}_k\times{\bf S}_i)^zS_j^z \Big] \Big\} \nonumber\\
 && + \sum_{k \in N(j)}\Big\{({\bf S}_i\times{\bf S}_k)^z S^z_j \nonumber\\
 && + \Delta\Big[ ({\bf S}_j\times{\bf S}_i)^zS_k^z  + ({\bf S}_k\times{\bf S}_j)^zS_i^z \Big] \Big\}\Big) \nonumber\\
&=& \frac{J^2}{4}\sum_i  \sum_{j,k \in N(i)} \big( {\bf r}_j - {\bf r}_k \big) \Big\{({\bf S}_j\times{\bf S}_k)^z S^z_i \nonumber\\
&& + \Delta \Big[ ({\bf S}_j \times {\bf S}_k)^x S^x_i + ({\bf S}_j \times {\bf S}_k) S^y_i \Big] \Big\},
\end{eqnarray} 
where Eq. (\ref{eq:Bloch}) has been used in going from the first line to the second line for each current. The obtained result is essentially the same as the previously obtained expressions \cite{SpinDyn_Huber_74, SpinDyn_Jencic_prb_15, MHall_Mook_prb_16, MHall_Mook_prb_17, Thermal_Huber_ptp_68, SpinDyn_Zotos_prb_05, SpinDyn_Sentef_07, SpinDyn_Pires_09, SpinDyn_Chen_13}. We note that Eqs. (\ref{eq:j_sc}) and (\ref{eq:j_th}) for the classical spin systems can also be applied for quantum spin systems by merely replacing ${S}_i^\alpha$ with the associated spin operator $\hat{S}_i^\alpha$. Indeed, one can verify that with the use of the Heisenberg equation instead of Eq. (\ref{eq:Bloch}), the same expressions as Eqs. (\ref{eq:j_sc}) and (\ref{eq:j_th}) are obtained.
 
\begin{figure}[t]
\includegraphics[width=\columnwidth]{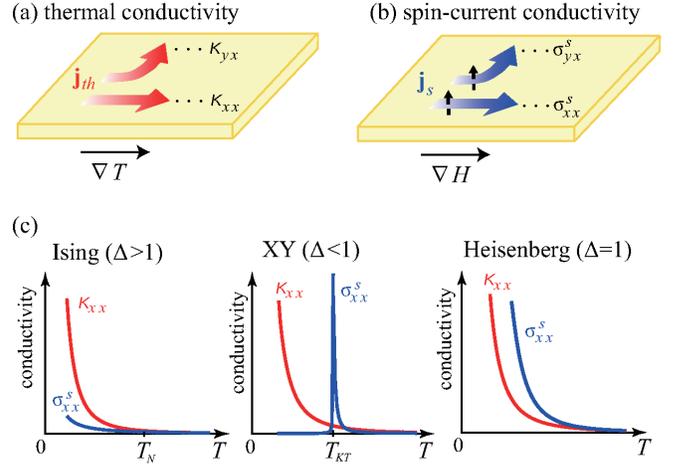}
\caption {System setups for the measurements of (a) thermal conductivity and (b) spin-current conductivity. In (b), the magnetic-anisotropy axis, which corresponds to the polarization direction of the spin current denoted by a black arrow, is assumed to be perpendicular to the two-dimensional sample plane. (c) Summary of our result: schematic temperature dependences of the longitudinal thermal conductivity $\kappa_{xx}$ (red curves) and spin-current conductivity $\sigma^s_{xx}$ (blue curves) in the Ising-type ($\Delta <1$), $XY$-type ($\Delta<1$), and Heisenberg-type ($\Delta=1$) spin systems in the thermodynamic limit. In contrast to $\kappa_{xx}$ commonly following a power-law behavior at low temperatures, $\sigma^s_{xx}$ exhibits temperature dependences characteristic of the three different universality classes. In particular, in the $XY$ case, $\sigma^s_{xx}$ exhibits a divergent sharp peak at the KT transition temperature $T_{KT}$. \label{fig:overview}}
\end{figure}

Next, we turn to the conductivities of the spin and thermal currents. We first introduce the theoretical framework for the quantum mechanical systems, and then, take the classical limit. 
In general, driving forces for the spin and thermal currents are magnetic-field and temperature gradients, $\nabla H$ and $\nabla T$, respectively [see Figs. \ref{fig:overview} (a) and (b)], so that the linear response equations are given by
\begin{equation}
\left( \begin{array}{c}
{\bf j}_s \\
{\bf j}_{th} 
\end{array} \right) = \left( \begin{array}{cc}
L^{s,s} & L^{s,th} \\
L^{th,s} & L^{th,th} 
\end{array}\right)\left(\begin{array}{c}
\nabla H \\
\nabla T/T 
\end{array}\right)
\end{equation}  
with the spin and thermal current-densities ${\bf j}_s$ and ${\bf j}_{th}$ \cite{MHall_Mook_prb_16, MHall_Mook_prb_17, book_Mahan}. Then, the spin-current conductivity $\sigma^s$ and the thermal conductivity $\kappa$ are expressed as
\begin{equation}
\sigma^s=L^{s,s}, \quad \kappa=T^{-1}L^{th,th}.
\end{equation} 
Note that in the present model without a magnetic field, $L^{s,th}=L^{th,s}=0$ is satisfied because these quantities are odd with respect to spins. 
In the linear response theory \cite{KuboFormular_Kubo_57}, the coefficients $L^{a,b}$ can be calculated from the formula
\begin{equation}\label{eq:conductivity_quantum_original}
L^{a,b}_{\mu\nu}(\omega) = \int_0^\infty dt \, e^{-i\omega t -\eta t} \int_0^{1/T} d\lambda \, \big\langle j_{a,\nu}(-i\hbar \lambda)j_{b,\mu}(t) \big\rangle,
\end{equation}
where $\langle ... \rangle$ denotes the thermal average in the equilibrium state.
Now, we will take the classical limit of Eq. (\ref{eq:conductivity_quantum_original}).
In the classical system, by making $\hbar \rightarrow 0$ \cite{KuboFormular_Kubo_57}, we have
\begin{equation}
L^{a,b}_{\mu\nu}(0) = \frac{1}{T} \int_0^\infty dt \, \big\langle j_{a,\nu}(0) \, j_{b,\mu}(t) \big\rangle.
\end{equation}
Thus, in the present classical XXZ model, we obtain the following expressions for the spin-current and thermal conductivities: 
\begin{eqnarray}\label{eq:conductivity}
\sigma_{\mu \nu}^s &=& \frac{1}{T \, L^2} \int_0^\infty dt \, \big\langle J^z_{s,\nu}(0) \, J^z_{s,\mu}(t) \big\rangle, \nonumber\\
\kappa_{\mu \nu} &=& \frac{1}{T^2 \, L^2} \int_0^\infty dt \, \big\langle J_{th,\nu}(0) \, J_{th,\mu}(t) \big\rangle,
\end{eqnarray} 
where we have used the relation between the total current and its current density, ${\bf j}_s={\bf J}^z_s/L$ and ${\bf j}_{th}={\bf J}_{th}/L$, with $L$ being a linear system size \cite{SpinDyn_Zotos_prb_05, SpinDyn_Jencic_prb_15, MHall_Mook_prb_16, MHall_Mook_prb_17}. Now, the problem is reduced to calculate the time correlations of the spin and thermal currents at various temperatures. In the present square lattice, the total number of spin $N_{\rm spin}$ and the system size $L$ are related by $L^2=N_{\rm spin} \, a^2$, where $a$ is a lattice constant. Noting that the time $t$ is measured in units of $|J|^{-1}$, it turns out that $\sigma^s_{\mu \nu}$ is a dimensionless quantity and $\kappa_{\mu \nu}$ has the dimension of $|J|$. Although in Eqs. (\ref{eq:j_sc}) and (\ref{eq:j_th}), the currents themselves involve the dimension of length, the conductivities in the present two-dimensional system do not, so that the length scale of the lattice constant $a$ is not relevant and thus, we take $a=1$ throughout this paper except for the case where $a$ is explicitly written. 

\subsection{Numerical method}
The time evolutions of ${\bf J}^z_s$ and ${\bf J}_{th}$ are determined microscopically by the spin-dynamics equation (\ref{eq:Bloch}), so that we numerically integrate Eq. (\ref{eq:Bloch}) and calculate the time correlations $\langle J^z_{s,\nu}(0) \, J^z_{s,\mu}(t) \rangle$ and $\langle J_{th,\nu}(0) \, J_{th,\mu}(t) \rangle$ at each time step. In the numerical integration of Eq. (\ref{eq:Bloch}), we use the second order symplectic method which guarantees the exact energy conservation \cite{Symplectic_Krech_98, Sqomega_Okubo_jpsj_10, Symplectic_Furuya_11}. We have confirmed that numerical results shown below are not altered if the 4th order Runge-Kutta method is used instead of the symplectic method. To properly evaluate the integral over time in Eq. (\ref{eq:conductivity}), we perform long-time integrations typically up to $t=100\,|J|^{-1} \, - \, 800 \, |J|^{-1}$ with the time step $\delta t=0.01 \, |J|^{-1}$ until the time correlations $\big\langle J^z_{s,\nu}(0) \, J^z_{s,\mu}(t) \big\rangle$ and $\big\langle J_{th,\nu}(0) \, J_{th,\mu}(t) \big\rangle$ are completely lost.  

Since Eq. (\ref{eq:Bloch}) does not have a phenomenological dissipation term, the thermal fluctuations are only one possible cause for the current relaxation. Although Eq. (\ref{eq:Bloch}) itself is deterministic, such a temperature effect can be incorporated by using temperature-dependent equilibrium spin configurations as the initial states for the equation of motion (\ref{eq:Bloch}). In order to thermalize the system to given temperature $T$, we perform MC simulations for the spin Hamiltonian (\ref{eq:Hamiltonian}). The thermal average is taken as the average over initial equilibrium spin configurations generated in the MC simulations.    
In this work, at each temperature, we prepared 2000-4000 equilibrium spin configurations by picking up a spin snapshot in every 1000 MC sweeps after 10$^5$ MC sweeps for thermalization, where one MC sweep consists of the 1 heat-bath sweep and successive 10-30 over-relaxation sweeps. 

By carefully analyzing the system-size dependences of the spin-current conductivity $\sigma^s_{\mu \nu}$ and the thermal conductivity $\kappa_{\mu \nu}$ at given temperatures, we will discuss the temperature dependences of $\sigma^s_{\mu \nu}$ and $\kappa_{\mu \nu}$ in the thermodynamic limit ($L \rightarrow \infty$) of our interest. 

\section{Analytical results in the low-temperature limit: calculations based on the linear spin-wave theory}
Before discussing numerical results, we should know how $\kappa_{\mu \nu}$ and $\sigma^s_{\mu \nu}$ should behave in the low-temperature limit. In this section, we will analytically investigate the temperature dependences of $\kappa_{\mu\nu}$ and $\sigma^s_{\mu \nu}$ based on the linear spin-wave theory (LSWT). As a low-temperature ordered state is a starting point in LSWT, one might be afraid that LSWT cannot be applied to the Heisenberg case because of the absence of the long-range order at any finite temperature. As long as there is a long-range order at $T=0$, however, the spin-wave expansions could still be done locally within the regions smaller than the spin-correlation length $\xi_s$ \cite{MagnonDamping_Tyc_89}. Thus, in the Heisenberg case, we introduce a lower cutoff in the momentum space which corresponds to the inverse spin-correlation length $\xi_s^{-1}$, and take the temperature dependence of $\xi_s \sim a \exp[b_H |J|/T]$ into account, where $b_H \simeq 2\pi$ is a universal constant \cite{Heisenberg_Polyakov_75}. 

In this section, we will start from the theory of the corresponding quantum spin system, and then, take the classical limit of relevant physical quantities.
By performing the spin-wave expansion, one can obtain the magnon representation of the Hamiltoninan (\ref{eq:Hamiltonian}) and the spin and thermal currents in Eqs.  (\ref{eq:j_sc}) and (\ref{eq:j_th}). 
Since in Eq. (\ref{eq:conductivity}), the time correlation functions $\langle J_{th,\nu}(0) \, J_{th,\mu}(t) \rangle$ and $\langle J^z_{s,\nu}(0) \, J^z_{s,\mu}(t) \rangle$ are essential for $\kappa_{\mu \nu}$ and $\sigma^s_{\mu \nu}$, we will first examine the associated static thermodynamic quantities, the equal-time correlation function $\langle J_{th,\nu}(0) \, J_{th,\mu}(0) \rangle$ and $\langle J^z_{s,\nu}(0) \, J^z_{s,\mu}(0) \rangle$. Then, the dynamical quantities, i.e., $\kappa_{\mu \nu}$ and $\sigma^s_{\mu \nu}$ due to the magnon propagation, will be calculated, putting emphasis on their temperature dependences in the classical limit.    
As we will see below, the temperature dependence of the thermal conductivity $\kappa_{\mu \nu}$ is almost independent of the magnetic anisotropy $\Delta$, while the spin-current conductivity $\sigma^s_{\mu \nu}$ is sensitive to the ordering properties controlled by $\Delta$. 
 
\subsection{Magnon representation}
Although our target system in the present paper is the classical XXZ model, we consider, for convenience, the corresponding quantum spin system throughout this subsection.   
The magnon representation of the Hamiltonian (\ref{eq:Hamiltonian}) and the spin and thermal currents in Eqs. (\ref{eq:j_sc}) and (\ref{eq:j_th}) can be derived by using the spin-wave expansions. In the Ising case of $\Delta > 1$, the quantization axis of spin is in the $z$ direction, so that we introduce the transformation from the laboratory frame to the rotated frame with $y$ being the rotation axis, 
\begin{equation}
\left\{\begin{array}{l} 
S^x_i= \tilde{S}^z_i \sin(\theta_i) + \tilde{S}^x_i \cos(\theta_i) \\
S^z_i= \tilde{S}^z_i \cos(\theta_i) - \tilde{S}^x_i \sin(\theta_i) \\
S^y_i= \tilde{S}^y_i 
\end{array} \right . ,\nonumber
\end{equation} 
where $\theta_i = {\bf Q}\cdot {\bf r}_i$ and ${\bf Q}=(\pi,\pi)$ is the ordering vector of the two-sublattice antiferromagnetic order. Then, the Hamiltonian reads
\begin{equation}
{\cal H} = \frac{J}{2} \sum_ i \sum_{j \in N(i)} \Big[ \tilde{S}^x_i \tilde{S}^x_j - \tilde{S}^y_i \tilde{S}^y_j + \Delta \tilde{S}^z_i \tilde{S}^z_j     \Big].
\end{equation} 
By using the Holstein-Primakoff transformation
\begin{equation}
\left\{\begin{array}{l} 
\tilde{S}^z_i = S- \hat{a}^\dagger_i \hat{a}_i \\
\tilde{S}^x_i + i \tilde{S}^y_i = \sqrt{2S}\Big(1-\frac{\hat{a}^\dagger_i \hat{a}_i}{2S} \Big)^{\frac{1}{2}}\hat{a}_i = \sqrt{2S} \, \hat{a}_i +{\cal O}(S^{-\frac{1}{2}})\\
\tilde{S}^x_i - i \tilde{S}^y_i = \sqrt{2S}\hat{a}^\dagger_i\Big(1-\frac{\hat{a}^\dagger_i \hat{a}_i}{2S} \Big)^{\frac{1}{2}} = \sqrt{2S} \, \hat{a}^\dagger_i+{\cal O}(S^{-\frac{1}{2}}) \\
\end{array} \right . 
\end{equation}
with $\hat{a}^\dagger_i$ and $\hat{a}_i$ being respectively the bosonic creation and annihilation operators and the Fourier transformation of these operators 
\begin{equation}
\hat{a}^\dagger_i = \frac{1}{\sqrt{N}}\sum_{\bf q} \hat{a}^\dagger_{\bf q} e^{-i{\bf q}\cdot {\bf r}_i}, \quad \hat{a}_i = \frac{1}{\sqrt{N}}\sum_{\bf q} \hat{a}_{\bf q} e^{i{\bf q}\cdot {\bf r}_i},
\end{equation}
we obtain
\begin{eqnarray}
{\cal H} &=&\frac{1}{2}\sum_{\bf q}\Big[ A_{\bf q} \big( \hat{a}^\dagger_{\bf q}\hat{a}_{\bf q} + \hat{a}_{\bf q}\hat{a}^\dagger_{\bf q} \big)- B_{\bf q} \big(\hat{a}^\dagger_{\bf q}\hat{a}^\dagger_{-{\bf q}}+ \hat{a}_{\bf q}\hat{a}_{-{\bf q}} \big) \Big] \nonumber\\
&+& const. + {\cal O}(S^0), \nonumber
\end{eqnarray}
where $A_{\bf q} = -4JS\Delta$, $B_{\bf q} = -4JS \gamma_{\bf q}$, and $\gamma_{\bf q} = \frac{1}{2}\big[ \cos(q_x)+\cos(q_y) \big]$.
The above Hamiltonian for the $\hat{a}_{\bf q}$ magnons can be diagonalized with the help of the Bogoliubov transformation
\begin{equation}
\left\{\begin{array}{l} 
\hat{a}_{\bf q} = u_{\bf q} \, \hat{b}_{\bf q} + v_{\bf q} \, \hat{b}^\dagger_{-{\bf q}}, \\
u_{\bf q}=u_{-{\bf q}} = \frac{1}{2}\Big[ \Big(\frac{A_{\bf q}+B_{\bf q}}{A_{\bf q}-B_{\bf q}} \Big)^{1/4} + \Big(\frac{A_{\bf q}-B_{\bf q}}{A_{\bf q}+B_{\bf q}} \Big)^{1/4}\Big], \nonumber\\
v_{\bf q}=v_{-{\bf q}} = \frac{1}{2}\Big[ \Big(\frac{A_{\bf q}+B_{\bf q}}{A_{\bf q}-B_{\bf q}} \Big)^{1/4} - \Big(\frac{A_{\bf q}-B_{\bf q}}{A_{\bf q}+B_{\bf q}} \Big)^{1/4}\Big],
\end{array} \right .  
\end{equation}
where $\hat{b}^\dagger_{\bf q}$ and $\hat{b}_{\bf q}$ are the creation and annihilation operators for magnons. 
In the $XY$ ($\Delta <1$) and the Heisenberg ($\Delta=1$) cases, we take the quantization axis in the $x$ and $z$ directions, respectively. The diagonalized magnon Hamiltonian in the three cases, $\Delta>1$, $\Delta=1$, and $\Delta<1$, is summarized as follows:
\begin{eqnarray}\label{eq:Hamiltonian_mag}
{\cal H} &\simeq& \sum_{\bf q} \varepsilon_{\bf q} \, \hat{b}^\dagger_{\bf q}\hat{b}_{\bf q}, \qquad \varepsilon_{\bf q} = \sqrt{A_{\bf q}^2-B_{\bf q}^2}, \nonumber\\
A_{\bf q} &=& 4|J|S \left\{\begin{array}{l} 
\Delta \qquad\qquad\qquad\quad (\Delta \geq 1) \\
1 - \frac{1}{2} (1-\Delta)\gamma_{{\bf q}} \quad (\Delta < 1) \\
\end{array} \right . \nonumber\\
B_{\bf q} &=& 4|J|S \left\{\begin{array}{l}
\gamma_{\bf q} \qquad\qquad\quad (\Delta \geq 1) \\
\frac{1}{2}(1+\Delta)\gamma_{{\bf q}} \quad (\Delta < 1) \\
\end{array} \right . \nonumber\\
\gamma_{\bf q} &=& \frac{1}{2}\big[ \cos(q_x a)+\cos(q_y a) \big] ,
\end{eqnarray} 
where we have dropped constant and higher-order terms.
Note that in the $XY$ and Heisenberg cases of $\Delta\leq 1$, the magnon is a gapless excitation, while in the Ising case of $\Delta > 1$, the magnon excitation has the gap $\Delta_{gp}=4|J|S\sqrt{\Delta^2-1}$. 

In the same manner, the thermal and spin currents in Eqs. (\ref{eq:j_th}) and (\ref{eq:j_sc}) can be expressed by the $\hat{b}_{\bf q}$ magnons as follows:
\begin{equation}\label{eq:current_th_mag}
{\bf J}_{th} = \big( 4|J|S\big)^2 \sum_{\bf q} \tilde{\varepsilon}_{\bf q} \, \tilde{{\bf v}}_{\bf q}  \, \hat{b}_{\bf q}^\dagger \hat{b}_{\bf q} + {\cal O}\big( S^1 \big) ,
\end{equation}
\begin{equation}\label{eq:current_spin_mag}
{\bf J}^z_s = \left\{\begin{array}{l} 
2 |J| S   \sum_{\bf q} \tilde{{\bf v}}_{\bf q}  \Big[  \frac{A_{\bf q}}{B_{\bf q}} ( \hat{b}_{\bf q}^\dagger \hat{b}_{-{\bf q}+{\bf Q}}^\dagger + \hat{b}_{\bf q} \hat{b}_{-{\bf q}+{\bf Q}} )\\
\qquad -( \hat{b}_{\bf q}^\dagger \hat{b}_{{\bf q}+{\bf Q}} + \hat{b}_{\bf q} \hat{b}_{{\bf q}+{\bf Q}}^\dagger )    \Big] + {\cal O}\big( S^0 \big)   \quad (\Delta \geq 1) \\
{\cal O}\big( S^{1/2} \big) \qquad\qquad\qquad\qquad\qquad\qquad\quad (\Delta < 1), \\
\end{array} \right . 
\end{equation}
where
\begin{eqnarray}
\tilde{\varepsilon}_{\bf q} &=& \frac{\varepsilon_{\bf q}}{4|J|S} \nonumber\\
\tilde{{\bf v}}_{\bf q} &=& \frac{{\bf v}_{\bf q}}{4|J|S}, \quad {\bf v}_{\bf q}=\nabla_{\bf q} \varepsilon_{\bf q}.
\end{eqnarray}
Since ${\bf v}_{\bf q}=\nabla_{\bf q} \varepsilon_{\bf q}$ represents the magnon velocity, the thermal current ${\bf J}_{th}$ can be regarded as the energy flow carried by the magnons. In contrast to the thermal current ${\bf J}_{th}$ having the common magnon-representation independent of the magnetic anisotropy $\Delta$, the spin current in the Ising and Heisenberg cases ($\Delta \geq 1$) is expressed in the form fundamentally different from the one in the $XY$ case ($\Delta<1$). The former has the leading order contribution of the order of ${\cal O}\big( S^1 \big)$, while the latter does not. In the $XY$ case, the spin current due to the magnon propagation is of the order of ${\cal O}\big( S^{1/2} \big)$. As the spin-wave expansion is the expansion with respect to $1/S$, such a higher order term is dropped in spirits of LSWT, so that ${\bf J}^z_s$ vanishes in the low-temperature ordered phase of the $XY$-type spin systems. The difference between $\Delta \geq 1$ and $\Delta<1$ cases stems from the difference in the direction of the quantization axis of spin: for $\Delta \geq 1$, the quantization axis is in the $z$ direction, whereas for $\Delta <1$, it is in the $xy$ plane which is perpendicular to the spin polarization of the spin current ${\bf J}^z_s$. Remember that although the spin current has its foundation on the conservation of the magnetization, only the $z$ component of the magnetization is conserved in the present anisotropic XXZ model in Eq. (\ref{eq:Hamiltonian}). 

\subsection{Static physical quantities}
As the magnon Hamiltonian (\ref{eq:Hamiltonian_mag}) is already diagonalized, one can easily calculate the thermal average of the current-related static quantities, $\langle J_{th,\nu}(0) \, J_{th,\mu}(0) \rangle$ and $\langle J^z_{s,\nu}(0) \, J^z_{s,\mu}(0) \rangle$. 

First, we consider the thermal average of the equal-time correlation function for the thermal current, $\langle J_{th,\nu}(0) \, J_{th,\mu}(0) \rangle$. With the use of the magnon representation in Eq. (\ref{eq:current_th_mag}), we have
\begin{eqnarray}\label{eq:Jth_static_tmp}
&&\big\langle J_{th,\nu}(0) \, J_{th,\mu}(0) \big\rangle = \sum_{{\bf q},{\bf q}'} \varepsilon_{\bf q} \varepsilon_{{\bf q}'} v_{{\bf q},\mu} v_{{\bf q}',\nu} \big\langle \hat{b}_{\bf q}^\dagger \hat{b}_{\bf q} \hat{b}_{{\bf q}'}^\dagger \hat{b}_{{\bf q}'} \big\rangle \nonumber\\
&& \qquad = \delta_{\mu,\nu} \sum_{\bf q} \big[ \varepsilon_{{\bf q}} \, v_{{\bf q},\mu} \big]^2 f_{\rm B}(\varepsilon_{\bf q})\big[1+2f_{\rm B}(\varepsilon_{\bf q}) \big],
\end{eqnarray}
where we have used the formula
\begin{eqnarray}
&& \big\langle \hat{b}_{\bf q}^\dagger \hat{b}_{\bf q} \hat{b}_{{\bf q}'}^\dagger \hat{b}_{{\bf q}'} \big\rangle = \frac{T^2}{Z} \frac{\partial ^2 \, Z}{\partial \varepsilon_{\bf q} \partial \varepsilon_{{\bf q}'}}, \\
&& Z = {\rm Tr} \Big[ \exp\big( - \frac{1}{T}\sum_{\bf q}\varepsilon_{\bf q} \hat{b}_{\bf q}^\dagger \hat{b}_{\bf q} \big) \Big] = \prod_{\bf q}\big[-f_{\rm B}(-\varepsilon_{\bf q}) \big] \nonumber
\end{eqnarray}
with the Bose-Einstein distribution function $f_{\rm B}(x)=(e^{x/T}-1)^{-1}$. Note that in Eq. (\ref{eq:Jth_static_tmp}), the off-diagonal term of $\mu \neq \nu$ vanishes after the summation over ${\bf q}$ because $v_{{\bf q},\mu} \propto \sin(q_{\mu})$ is an odd function of ${\bf q}$. 

Now, we shall move on to the classical spin system. In the classical limit of
\begin{equation}\label{eq:classical_limit}
f_{\rm B}(x) \rightarrow \frac{T}{x},
\end{equation}
the equal-time correlation for the classical spins $\langle J_{th,\nu}(0) \, J_{th,\mu}(0) \rangle_{\rm cl}$ is obtained as
\begin{equation}\label{eq:Jth_static}
\big\langle J_{th,\nu}(0) \, J_{th,\mu}(0) \big\rangle_{\rm cl} = \delta_{\mu,\nu} 2 \, T^2  \, \sum_{\bf q} \big[  v_{{\bf q},\mu} \big]^2.
\end{equation}
At this point, the $T^{2}$ dependence of $\langle J_{th,\nu}(0) \, J_{th,\mu}(0) \rangle_{\rm cl}$ is clear at least in the Ising and $XY$ cases. In the Heisenberg case, however, the additional temperature dependence due to the spin-correlation length $\xi_s$ comes in through the summation over ${\bf q}$. As we mentioned in the beginning of this section, $\xi_s$ enters in the form of the lower cutoff in the ${\bf q}$ space, i.e., $\xi_s^{-1} \leq |{\bf q}|$. For completeness, we shall evaluate the summation over ${\bf q}$ in Eq. (\ref{eq:Jth_static}) in all the three cases. Since the dominant contribution comes from the low-energy excitation near $|{\bf q}|\simeq 0$, we have
\begin{equation}\label{eq:magenergy_app}
\varepsilon_{\bf q} \simeq \left\{ \begin{array}{l}
\sqrt{\Delta_{gp}^2+ \frac{(4|J|S)^2}{2} |{\bf q}|^2} \qquad (\Delta > 1) \\
2\sqrt{2}|J|S \, |{\bf q}| \qquad\qquad\quad (\Delta = 1) \\
2|J|S\sqrt{ 1+\Delta } \, |{\bf q}| \qquad\quad (\Delta < 1) \\
\end{array} \right. . \nonumber 
\end{equation}
Then, the ${\bf q}$-summation can be replaced with the following integral over $\varepsilon_{\bf q}$,
\begin{eqnarray}\label{eq:qsum}
&& \sum_{\bf q} \simeq \frac{L^2}{(2\pi)^2} \int_0^{2\pi} d\phi_{\bf q} \int_{\varepsilon_{\rm min}}^{\varepsilon_{\rm max}} d\varepsilon_{\bf q} \, D(\varepsilon_{\bf q}), \nonumber\\
&& \varepsilon_{\rm min} = \left\{ \begin{array}{c}
\Delta_{gp} \qquad\quad (\Delta > 1) \\
\displaystyle{\frac{2\sqrt{2}|J|S}{\xi_s/a} \quad (\Delta = 1)} \\
0 \qquad\qquad (\Delta < 1) \\
\end{array} \right., 
\end{eqnarray} 
where the density of states $D(\varepsilon_{\bf q})$ and the higher energy cutoff $\varepsilon_{\rm max}$ are given by $D(\varepsilon_{\bf q})= 2 \varepsilon_{\bf q}/(4|J|S)^2$ and $\varepsilon_{\rm max} \sim 4|J|S \Delta $ for $\Delta \geq 1$, and $D(\varepsilon_{\bf q})= [4/(1+\Delta)] \varepsilon_{\bf q}/(4|J|S)^2$ and $\varepsilon_{\rm max} \sim 4|J|S $ for $\Delta < 1$. Note that in the Heisenberg case of $\Delta=1$, the low-energy cutoff $\varepsilon_{\rm min}$ possesses the temperature dependence via the spin-correlation length $\xi_s/a \sim \exp[b_H |J|/T]$. As we will see below, this additional temperature dependence coming from $\xi_s$ is negligibly small for the thermal transport, but not for the spin transport. By using Eq. (\ref{eq:qsum}) and performing the integral over $\varepsilon_{\bf q}$, we can evaluate the ${\bf q}$-summation in Eq. (\ref{eq:Jth_static}) to yield 
\begin{eqnarray}\label{eq:Jth_static_final}
&&\big\langle J_{th,\nu}(0) \, J_{th,\mu}(0) \big\rangle_{\rm cl} /L^2\simeq \delta_{\mu,\nu} \frac{T^2}{8 \pi} (4|J|S)^2 \nonumber\\
&&\times\left\{ \begin{array}{l}
\displaystyle{ 1-\Big(\frac{\Delta^2-1}{\Delta}\Big)^2+2\frac{\Delta^2-1}{\Delta}\ln\Big(\frac{\Delta^2-1}{\Delta}\Big)} \qquad (\Delta>1) \\
\displaystyle{ 1-\frac{1}{2}\Big(\frac{a}{\xi_s}\Big)^2 } \qquad (\Delta=1) \\
\displaystyle{ 2-\frac{4\Delta}{(1+\Delta)^2} } \qquad (\Delta<1) \\
\end{array} \right.
\end{eqnarray}
As the correction $(a/\xi_s)^2$ in the $\Delta=1$ case is negligibly small, $\langle J_{th,\nu}(0) \, J_{th,\mu}(0) \rangle$ in the classical limit exhibits the $T^2$ behavior at low temperatures, being independent of the magnetic anisotropy $\Delta$.
 
Next, we calculate the equal-time correlation function for the spin current in the classical limit, $\langle J^z_{s,\nu}(0) \, J^z_{s,\mu}(0) \rangle_{\rm cl}$. Since in the $XY$ case of $\Delta<1$, the spin current is absent within the leading-order magnon contribution [see Eq. (\ref{eq:current_spin_mag})], we only consider the $\Delta \geq 1$ case in which after some manipulations, we have
\begin{eqnarray}\label{eq:Js_static_pretmp}
&&\big\langle J^z_{s,\nu}(0) \, J^z_{s,\mu}(0) \big\rangle = \frac{-1}{4}\sum_{{\bf q},{\bf q}'} v_{{\bf q},\nu} v_{{\bf q}',\mu} \Big\{\big( \delta_{{\bf q},{\bf q}'} + \delta_{{\bf q},{\bf q}'+{\bf Q}}\big) \nonumber\\
&& \quad \times \big[ f_{\rm B}(\varepsilon_{\bf q})f_{\rm B}(-\varepsilon_{{\bf q}+{\bf Q}})+f_{\rm B}(-\varepsilon_{\bf q})f_{\rm B}(\varepsilon_{{\bf q}+{\bf Q}}) \big]\nonumber\\
&& \, -\frac{A_{\bf q}}{B_{\bf q}}\frac{A_{{\bf q}'}}{B_{{\bf q}'}}\big( \delta_{{\bf q},{\bf q}'} + \delta_{{\bf q},-{\bf q}'+{\bf Q}}\big) \nonumber\\
&&\quad \times \big[f_{\rm B}(\varepsilon_{\bf q})f_{\rm B}(\varepsilon_{-{\bf q}+{\bf Q}})+f_{\rm B}(-\varepsilon_{\bf q})f_{\rm B}(-\varepsilon_{-{\bf q}+{\bf Q}}) \big] \Big\}.
\end{eqnarray}
Note that Eq. (\ref{eq:Js_static_pretmp}) is obtained for the quantum spin system. Now, we take the classical limit of Eq. (\ref{eq:Js_static_pretmp}). As the relations, $A_{\pm{\bf q}+{\bf Q}}=A_{\bf q}$, $B_{\pm{\bf q}+{\bf Q}}=-B_{\bf q}$, and ${\bf v}_{\pm{\bf q}+{\bf Q}}=\pm {\bf v}_{\bf q}$, are satisfied for $\Delta\geq 1$, the classical limit Eq. (\ref{eq:classical_limit}) yields
\begin{equation}\label{eq:Js_static_tmp}
\big\langle J^z_{s,\nu}(0) \, J^z_{s,\mu}(0) \big\rangle_{\rm cl}=\delta_{\mu,\nu}\sum_{\bf q} \big[ v_{{\bf q},\mu}\big]^2 \Big( 1+\frac{A_{\bf q}^2}{B_{\bf q}^2}\Big)\frac{T^2}{\varepsilon_{\bf q}^2}.
\end{equation}
By using Eq. (\ref{eq:qsum}), one can evaluate the summation over ${\bf q}$ in Eq. (\ref{eq:Js_static_tmp}). The final result is summarized as follows:
\begin{eqnarray}\label{eq:Js_static}
&& \big\langle J^z_{s,\nu}(0) \, J^z_{s,\mu}(0) \big\rangle_{\rm cl}/L^2 \simeq \delta_{\mu,\nu} \frac{T^2}{4\pi} \nonumber\\
&&\times\left\{ \begin{array}{l}
\displaystyle{(3\Delta^2-1)\ln\Big( \frac{\Delta}{\sqrt{\Delta^2-1}}\Big)-\frac{3}{2}} \qquad (\Delta>1) \\
\displaystyle{ \ln(8)-\frac{1}{2} +2\ln\Big( \frac{\xi_s}{a}\Big) + \frac{1}{4}\frac{a^2}{\xi_s^2} } \qquad (\Delta=1) \\
\displaystyle{ 0 } \qquad (\Delta<1) \\
\end{array} \right. .
\end{eqnarray}
Note that in the $XY$ case of $\Delta<1$, $\big\langle J^z_{s,\nu}(0) \, J^z_{s,\mu}(0) \big\rangle_{\rm cl}$ is zero because the spin current is absent within the leading-order magnon contribution [see Eq. (\ref{eq:current_spin_mag})].
In the Ising case of $\Delta>1$, the equal-time correlation of the spin current $\langle J^z_{s,\nu}(0) \, J^z_{s,\mu}(0) \rangle_{\rm cl}$ has the same $T^2$ dependence as $\langle J_{th,\nu}(0) \, J_{th,\mu}(0) \rangle_{\rm cl}$. In the Heisenberg case of $\Delta=1$, on the other hand, $\langle J^z_{s,\nu}(0) \, J^z_{s,\mu}(0) \rangle_{\rm cl}$ includes a non-negligible correction term coming from the temperature-dependent $\xi_s$, i.e., $T^2\ln( \xi_s/a) \sim b_H |J|\, T$, and thus, takes the form of $\langle J^z_{s,\nu}(0) \, J^z_{s,\mu}(0) \rangle_{\rm cl}/L^2 \sim \delta_{\mu \nu} ( const \, T^2 + T )$. The correction term ($\propto T$) becomes the leading order contribution at lower temperatures, which is in sharp contrast to $\langle J_{th,\nu}(0) \, J_{th,\mu}(0) \rangle_{\rm cl}$ with the irrelevant correction terms [see Eq. (\ref{eq:Jth_static_final})]. As we will see below, such a situation is also the case for the dynamical quantities.

\subsection{Dynamical physical quantities}  
In the classical spin systems, the conductivities $\kappa_{\mu\nu}$ and $\sigma^s_{\mu\nu}$  are obtained from the time-correlation of the associated currents [see Eq. (\ref{eq:conductivity})]. Here, we consider the current dynamics brought by the magnon propagation in the presence of the magnon-magnon scatterings. In order to calculate the thermal average of the time correlation, it is convenient to start from the quantum mechanical system and take the classical limit of Eq. (\ref{eq:classical_limit}) afterwards. In the quantum mechanical system, the dynamical correlation function $L^{a,b}_{\mu\nu}(\omega)$ in Eq. (\ref{eq:conductivity_quantum_original}) can be expressed in the following form \cite{book_AGD}:
\begin{eqnarray}
L^{a,a}_{\mu\nu}(\omega) &=& -\frac{Q^{a,R}_{\mu\nu}(\omega)-Q^{a,R}_{\mu\nu}(0)}{i\omega}, \nonumber\\
Q^{a,R}_{\mu\nu}(\omega) &=& Q^a_{\mu\nu}(\omega + i 0), \\ 
Q^a_{\mu\nu}(i\omega_l) &=& -\frac{1}{L^2}\int_0^{1/T}\big\langle T_{\tau} J_{a,\mu}(\tau)J_{a,\nu}(0) \big\rangle \, e^{i \omega_n \, \tau} d\tau. \nonumber
\end{eqnarray} 
Here, $Q^a_{\mu\nu}(i\omega_l)$ is a response function and $\omega_n=2\pi n T$ is the bosonic Matsubara frequency. Then, the thermal conductivity $\kappa_{\mu\nu}$ and the spin-current conductivity $\sigma^s_{\mu\nu}$ are given by
\begin{eqnarray}\label{eq:conductivity_quantum}
\kappa_{\mu\nu} &=&\frac{1}{T} i \frac{d \, Q^{th,R}_{\mu\nu}(\omega)}{d \, \omega} \Big|_{\omega =0}, \nonumber\\
\sigma^s_{\mu\nu} &=& i \frac{d \, Q^{s,R}_{\mu\nu}(\omega)}{d \, \omega} \Big|_{\omega =0}.
\end{eqnarray}

We first calculate the thermal conductivity $\kappa_{\mu\nu}$. For the thermal current carried by the magnons in Eq. (\ref{eq:current_th_mag}), the response function $Q^{th}_{\mu\nu}(i\omega_n)$ is given by \cite{book_AGD}
\begin{eqnarray}
Q^{th}_{\mu\nu}(i\omega_n) &=& \frac{-1}{L^2}\sum_{\bf q} \varepsilon_{\bf q}^2 v_{{\bf q},\mu} \, v_{{\bf q},\nu} \, T\sum_{\omega_m}{\cal D}_{\bf q}(i\omega_m){\cal D}_{\bf q}(i\omega_m+i\omega_n) \nonumber\\
&=& \frac{-1}{L^2}\sum_{\bf q} \varepsilon_{\bf q}^2 v_{{\bf q},\mu} \,  v_{{\bf q},\nu} \int_{-\infty}^{\infty}\frac{dx}{2\pi i} \big[ {\cal D}^R_{\bf q}(x)-{\cal D}^A_{\bf q}(x) \big] \nonumber\\
&& \times  \big[ {\cal D}^R_{\bf q}(x+i\omega_n) + {\cal D}^A_{\bf q}(x-i\omega_n) \big] \, f_{\rm B}(x) ,
\end{eqnarray}
where ${\cal D}^R_{\bf q}(x)$ (${\cal D}^A_{\bf q}(x)=\big[{\cal D}^R_{\bf q}(x)\big]^*$) is the retarded (advanced) magnon Green's function obtained by analytic continuation $i\omega_m \rightarrow \omega + i0$ in the temperature Green's function ${\cal D}_{\bf q}(i\omega_m)$ defined by 
\begin{equation}
{\cal D}_{\bf q}(\tau) = -\big\langle T_{\tau} \hat{b}_{\bf q}(\tau)\hat{b}_{\bf q}^\dagger(0) \big\rangle = T\sum_{\omega_m} {\cal D}_{\bf q}(i\omega_m) \, e^{-i\omega_m \tau}.
\end{equation} 
With the use of Eq. (\ref{eq:conductivity_quantum}), the thermal conductivity in the quantum system is formally expressed as
\begin{equation}\label{eq:conductivity_quantum_th}
\kappa_{\mu\nu}=\frac{T^{-1}}{4\pi L^2}\int_{-\infty}^\infty dx \sum_{\bf q} \varepsilon_{\bf q}^2 \, v_{{\bf q},\mu} \, v_{{\bf q},\nu} f_{\rm B}'(x)\big[ {\cal D}^R_{\bf q}(x)-{\cal D}^A_{\bf q}(x) \big]^2.
\end{equation}
Here, the magnon Green's function ${\cal D}_{\bf q}^R(x)$ is given by
\begin{equation}\label{eq:Green_mag}
{\cal D}_{\bf q}^R(x) = \frac{1}{x-\varepsilon_{\bf q}+i \alpha \, x} = \big[ {\cal D}_{\bf q}^A(x) \big]^\ast,
\end{equation}
where the dimensionless coefficient $\alpha$ represents the magnon damping which corresponds to the Gilbert damping in the LLG equation \cite{MagnonGreen_Yamaguchi_17, MagnonTrans_Tatara_15}. In general, the damping $\alpha$ originates from the interactions associated with spins in solids, so that it may be brought not only by the magnon-magnon scatterings but also, for example, by magnon-phonon scatterings. In the present work, however, the starting point is the spin Hamiltonian (\ref{eq:Hamiltonian}) and no further assumption is made. Thus, $\alpha$ is of purely magnetic origin and brought by the magnon-magnon scatterings. Since the temperature dependence of $\alpha$ has already been calculated in the typical case of $\Delta=1$ \cite{MagnonDamping_Tyc_89, MagnonDamping_Harris_71}, we will skip the microscopic derivation of $\alpha$ in this paper.

In the classical spin system, the concrete expression of Eq. (\ref{eq:conductivity_quantum_th}) can straightforwardly be derived, as shown below.
Substituting Eq. (\ref{eq:Green_mag}) into Eq. (\ref{eq:conductivity_quantum_th}) and taking the classical limit of $f_{\rm B}'(x)=-T/x^2$, we obtain the following expression for the thermal conductivity in the classical spin systems $\kappa_{\mu \nu}^{\rm cl}$ as
\begin{equation}
\kappa_{\mu \nu}^{\rm cl} = \frac{1}{2L^2}\frac{1+\alpha^2}{\alpha}\sum_{\bf q}\frac{1}{\varepsilon_{\bf q}} \, v_{{\bf q},\mu} \, v_{{\bf q},\nu},
\end{equation}
where the equation
\begin{equation}\label{eq:integral}
\int_{-\infty}^\infty \frac{dx}{ \big[(x-\varepsilon_{\bf q})^2+(\alpha x)^2\big]^2 } = \frac{\pi}{2}\frac{1+\alpha^2}{\varepsilon_{\bf q}^3 \alpha^3}
\end{equation}
has been used. The summation over ${\bf q}$ can be evaluated in the same manner as that for the static physical quantities. With the use of Eq. (\ref{eq:magenergy_app}), we obtain
\begin{eqnarray}\label{eq:conductivity_classical_th}
&& \kappa_{\mu \nu}^{\rm cl} \simeq \delta_{\mu,\nu}\frac{1}{12 \pi}\frac{1+\alpha^2}{\alpha} 4|J|S \nonumber\\
&& \times \left\{ \begin{array}{l}
\displaystyle{ \big( \Delta-\sqrt{\Delta^2-1} \big)^3  \qquad\qquad\qquad (\Delta > 1) } \\
\displaystyle{ \Big( 1-\frac{\sqrt{2}}{2}\frac{a}{\xi_s} \Big)^2 \Big( 1+\frac{\sqrt{2}}{4}\frac{a}{\xi_s} \Big) \qquad (\Delta =1) }  \\
\displaystyle{ \frac{3}{2}-\frac{2 \Delta}{\big( 1+\Delta \big)^2}  \qquad \qquad\qquad\quad (\Delta < 1) } \\
\end{array} \right. .
\end{eqnarray}
Only the longitudinal components of the thermal conductivity $\kappa_{\mu \mu}^{\rm cl}$ are non-vanishing. When the magnon damping is sufficiently small such that $\alpha \ll 1$, it follows that $\kappa_{\mu\nu}^{\rm cl}\propto 1/\alpha$, which agrees with the results obtained in other theoretical approaches \cite{MagnonTrans_Tatara_15, MagnonTrans_Jiang_13}.

One can see from Eq. (\ref{eq:conductivity_classical_th}) that in the Heisenberg case of $\Delta=1$, although the spin-correlation length $\xi_s$ rapidly increases toward $T=0$, such a temperature effect is irrelevant at lower temperatures because $\xi_s$ enters in $\kappa_{\mu\mu}^{\rm cl}$ in the form of $1/\xi_s$. Thus, in all the three ($\Delta > 1$, $\Delta=1$, and $\Delta<1$) cases, the temperature dependence of $\kappa_{\mu\mu}^{\rm cl} \propto 1/\alpha$ is governed by the magnon damping factor $\alpha$.

The damping of the antiferromagnetic magnon due to multi-magnon scatterings has already been calculated by using Feynman diagram techniques in Refs. \cite{MagnonDamping_Tyc_89, MagnonDamping_Harris_71}. The temperature dependence of $\alpha$ in the classical Heisenberg antiferromagnet essentially follows the $T^2$ form, i.e., $\alpha \propto T^2$, which results from the leading-order scattering process involving four magnons. In the $XY$-type and Ising-type classical spin systems, although the concrete expression of $\alpha$ is not available, the same temperature dependence $\alpha \propto T^2$ is expected because the same types of the Feynman diagrams (the same leading-order scattering processes) contribute to the magnon damping. Of course, there must be quantitative differences among the three cases. In particular, for the Ising-type anisotropy of $\Delta>1$, the magnon excitation is gapped, so that the phase space satisfying the energy conservation in the calculation of the relevant Feynman diagrams would be shrunk with increasing $\Delta$, resulting in a smaller value of $\alpha$. Apart from such a quantitative difference which may become serious for strong Ising-type anisotropies, the longitudinal thermal conductivity $\kappa_{\mu\mu}^{\rm cl}$ in the classical limit should behave as $\kappa_{\mu\mu}^{\rm cl} \propto 1/\alpha \propto 1/T^2$ in all the three ($\Delta > 1$, $\Delta=1$, and $\Delta<1$) cases.       

Now, we will move on to the calculation of the spin-current conductivity $\sigma^s_{\mu\nu}$ based on Eq. (\ref{eq:conductivity_quantum}). As in the case of the thermal current, starting from the magnon representation of the spin current in Eq. (\ref{eq:current_spin_mag}), we can write down the response function $Q^s_{\mu\nu}(i\omega_n)$ as
\begin{eqnarray}
&& Q^s_{\mu\nu}(i\omega_n) = \frac{-1}{4L^2}\sum_{{\bf q},{\bf q}'} v_{{\bf q},\mu} \, v_{{\bf q}',\nu} \, \Big\{ \big(\delta_{{\bf q},{\bf q}'}+\delta_{{\bf q},{\bf q}'+{\bf Q}} \big) F^+_{\bf q}(i \omega_n) \nonumber\\
&&\qquad\qquad\quad + \frac{A_{\bf q}}{B_{\bf q}}\frac{A_{{\bf q}'}}{B_{{\bf q}'}}\big(\delta_{{\bf q},{\bf q}'}+\delta_{{\bf q},-{\bf q}'+{\bf Q}} \big)F^-_{\bf q}(i \omega_n) \Big\} , \nonumber\\
&& F^\pm_{\bf q}(i \omega_n) = T\sum_{\omega_m} {\cal D}_{\bf q}(i\omega_m) \big[ {\cal D}_{{\bf Q}\pm {\bf q}}(i\omega_n\pm i\omega_m) \nonumber\\
&& \qquad\qquad\qquad\qquad\qquad\qquad + {\cal D}_{{\bf Q}\pm{\bf q}}(-i\omega_n\pm i\omega_m) \big] \nonumber\\
&& = \int_{-\infty}^\infty \frac{dx}{2\pi i} f_{\rm B}(x)\Big\{ \big[ {\cal D}^R_{{\bf Q}\pm {\bf q}}(\pm x+i \omega_n) + {\cal D}^A_{{\bf Q}\pm {\bf q}}(\pm x-i \omega_n) \big] \nonumber\\
&& \times \big[ {\cal D}^R_{\bf q}(x)- {\cal D}^A_{\bf q}(x) \big]  \pm \big[ {\cal D}^R_{{\bf Q}\pm {\bf q}}(\pm x) - {\cal D}^A_{{\bf Q}\pm {\bf q}}(\pm x) \big] \nonumber\\
&& \times \big[ {\cal D}^R_{\bf q}(x+i\omega_n) + {\cal D}^A_{\bf q}(x-i\omega_n) \big]  \Big\}. 
\end{eqnarray}
Then, the spin-current conductivity $\sigma^s_{\mu \nu}$ is formally written as
\begin{eqnarray}\label{eq:conductivity_quantum_spin}
&&\sigma^s_{\mu\nu} = \frac{1}{8\pi L^2 }\int_{-\infty}^\infty dx \sum_{{\bf q},{\bf q}'} v_{{\bf q},\mu} \, v_{{\bf q}',\nu} \, f_{\rm B}'(x) \big[ {\cal D}^R_{\bf q}(x)- {\cal D}^A_{\bf q}(x) \big] \nonumber\\
&&\times \Big\{ \big(\delta_{{\bf q},{\bf q}'}+\delta_{{\bf q},{\bf q}'+{\bf Q}} \big)  \big[ {\cal D}^R_{{\bf q}+{\bf Q}}(x)- {\cal D}^A_{{\bf q}+{\bf Q}}(x) \big] \\
&&-\frac{A_{\bf q}}{B_{\bf q}}\frac{A_{{\bf q}'}}{B_{{\bf q}'}} \big(\delta_{{\bf q},{\bf q}'}+\delta_{{\bf q},-{\bf q}'+{\bf Q}} \big) \big[ {\cal D}^R_{-{\bf q}+{\bf Q}}(x)- {\cal D}^A_{-{\bf q}+{\bf Q}}(x) \big] \Big\}. \nonumber
\end{eqnarray}
In the same manner as that for $\kappa_{\mu\nu}$, we take the classical limit of Eq. (\ref{eq:conductivity_quantum_spin}).
By substituting Eq. (\ref{eq:Green_mag}) into Eq. (\ref{eq:conductivity_quantum_spin}), taking the classical limit of $f_{\rm B}'(x)=-T/x^2$, and using Eq. (\ref{eq:integral}) and the formula
\begin{equation}
\int_{-\infty}^\infty \frac{dx}{ \big[(x-\varepsilon_{\bf q})^2+(\alpha x)^2\big]\big[(x+\varepsilon_{\bf q})^2+(\alpha x)^2\big] } = \frac{\pi}{2}\frac{1}{\varepsilon_{\bf q}^3 \alpha}, \nonumber
\end{equation}
we have the spin-current conductivity in the classical spin systems $\sigma^{s,{\rm cl}}_{\mu \nu}$ as follows:
\begin{equation}
\sigma^{s,{\rm cl}}_{\mu\nu} = \frac{1}{2L^2}T\sum_{\bf q} v_{{\bf q},\mu} \, v_{{\bf q},\nu} \frac{1}{\varepsilon_{\bf q}^3}\Big[ \frac{1+\alpha^2}{\alpha}+ \alpha \frac{A_{\bf q}^2}{B_{\bf q}^2} \Big]. 
\end{equation}
By further using the approximation Eq. (\ref{eq:qsum}), we finally obtain
\begin{eqnarray}\label{eq:conductivity_classical_spin}
&& \sigma^{s,{\rm cl}}_{\mu\nu} \simeq \delta_{\mu,\nu}\frac{1}{8 \pi} \frac{T}{4|J|S} \\
&& \times \left\{ \begin{array}{l}
\displaystyle{ \frac{\Delta-\sqrt{\Delta^2-1}}{\sqrt{\Delta^2-1}} \Big[ -\frac{2+\alpha^2}{3\alpha}\frac{1}{\Delta}} \\
\qquad\qquad\qquad \displaystyle{ + \frac{4+5\alpha^2}{3\alpha}\big( \Delta-\sqrt{\Delta^2-1}\big)  \Big]  \quad (\Delta > 1) } \\
\displaystyle{ \frac{1+2\alpha^2}{\alpha}\sqrt{2} \, \frac{\xi_s}{a} -\frac{2+3\alpha^2}{\alpha} +\frac{1+\alpha^2}{\alpha} \frac{\sqrt{2}}{2} \frac{a}{\xi_s}  \quad (\Delta =1) }  \nonumber\\
\displaystyle{ 0  \quad\quad (\Delta <1) } \nonumber\\
\end{array} \right. .
\end{eqnarray}
In contrast to the thermal conductivity $\kappa_{\mu\nu}^{\rm cl}$, the spin-current conductivity $\sigma^{s,{\rm cl}}_{\mu \nu}$ reflects the difference in the ordering properties. First of all, in the $XY$ case of $\Delta<1$, $\sigma^{s, {\rm cl}}_{\mu\nu}$ is zero because the spin current is absent within the leading-order magnon contribution [see Eq. (\ref{eq:current_spin_mag})]. In the Ising case of $\Delta>1$, as one can see from Eq. (\ref{eq:conductivity_classical_spin}), the temperature dependence of $\sigma^{s, {\rm cl}}_{\mu\mu}$ is determined by that of $T/\alpha$. Since for relatively weak anisotropies, $\alpha \propto T^2$ is expected to be satisfied, the longitudinal spin-current conductivity should exhibit the following temperature dependence: $\sigma^{s,{\rm cl}}_{\mu\mu} \propto T/\alpha \propto T^{-1}$. In the Heisenberg case of $\Delta=1$, one can see from Eq. (\ref{eq:conductivity_classical_spin}) that the spin-correlation length $\xi_s$ enters in the form of $\xi_s/\alpha$, so that the longitudinal spin-current conductivity should diverge toward $T=0$ in the exponential form of $\sigma^{s, {\rm cl}}_{\mu\mu} \propto \xi_s \, T/\alpha \sim \exp[b_H |J|/T]$. 

\begin{figure}[t]
\includegraphics[width=\columnwidth]{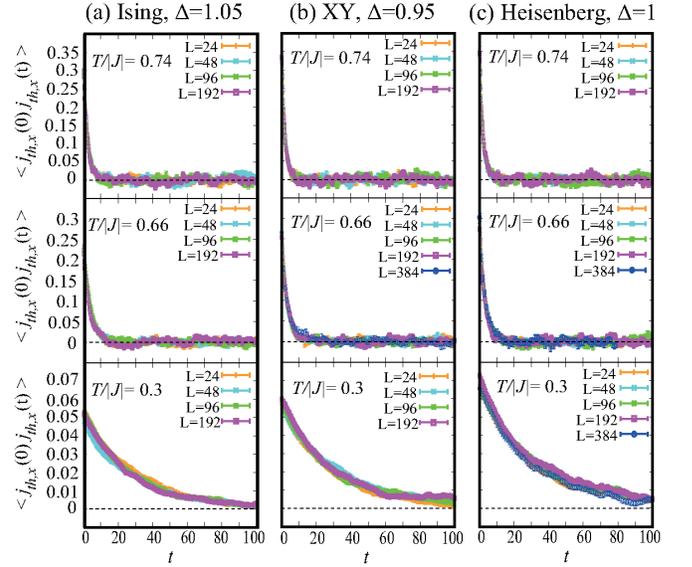}
\caption {The time correlation function of the thermal current $\langle j_{th,x}(0) \, j_{th,x}(t) \rangle$ at $T/|J|=0.74$ (top), $T/|J|=0.66$ (middle), and $T/|J|=0.3$ (bottom) in the (a) Ising-type ($\Delta=1.05$), (b) $XY$-type ($\Delta=0.95$), and (c) Heisenberg-type ($\Delta=1$) spin systems. Time $t$ and $\langle j_{th,x}(0) \, j_{th,x}(t) \rangle$ are measured in units of $|J|^{-1}$ and $|J|^4$, respectively. \label{fig:timedep_thermal_comp}}
\end{figure}

\begin{figure*}[t]
\includegraphics[scale=0.78]{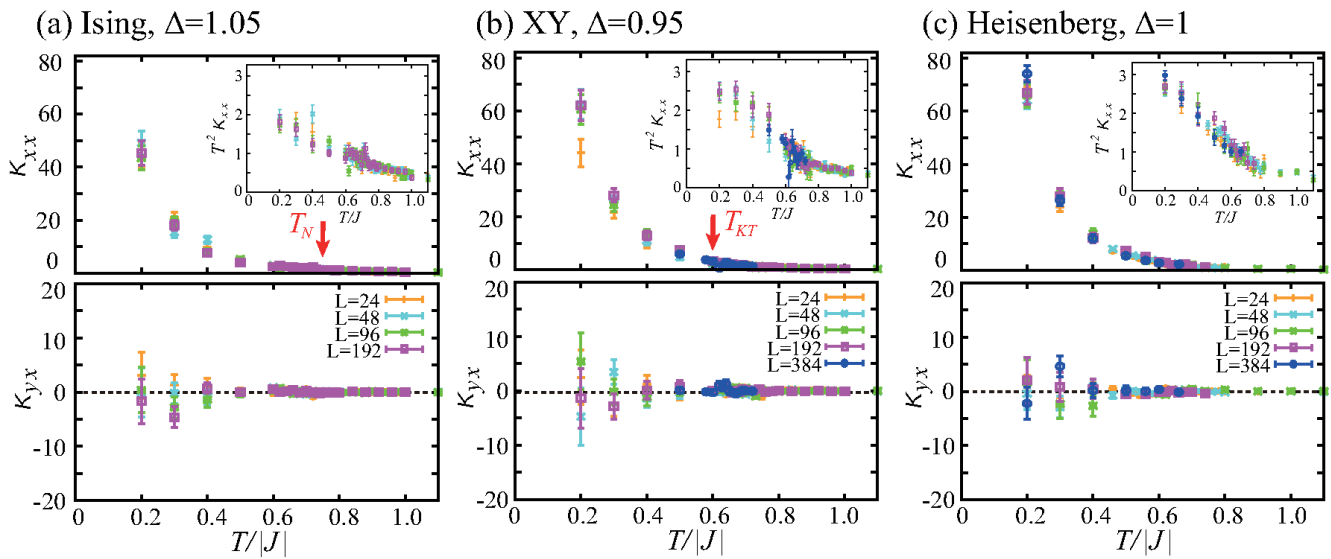}
\caption{The temperature dependence of the thermal conductivity $\kappa_{\mu \nu}$ in the (a) Ising-type ($\Delta=1.05$), (b) $XY$-type ($\Delta=0.95$), and (c) Heisenberg-type ($\Delta=1$) spin systems, where upper and lower panels show the longitudinal and transverse conductivities, respectively. $\kappa_{\mu \nu}$ is measured in units of $|J|$. In (a) and (b), red arrows indicate the magnetic and KT transition temperatures, $T_N/|J|\simeq 0.75$ and $T_{KT}/|J| \simeq 0.6$, respectively. Insets show $T^2 \, \kappa_{\mu\nu}$ in the same temperature range as that of the main panels. \label{fig:thermal_comp}}
\end{figure*}

In the following sections, we will show numerical results on $\kappa_{\mu\nu}$ and $\sigma^s_{\mu\nu}$, the low-temperature properties of which are qualitatively consistent with the above analytical results. It should be noted that the transport properties near the phase transition, which is our main focus of the present work, is out of the applicability range of LSWT.  

\section{Numerical results on the thermal conductivity}
In this section, we will discuss the association between the phase transition and the thermal transport based on numerical results obtained in the Ising-type ($\Delta > 1$), $XY$-type ($\Delta < 1$), and Heisenberg-type ($\Delta=1$) spin systems. In this paper, the parameter values of $\Delta=1.05$ and $\Delta=0.95$ are basically used in the Ising and $XY$ cases, respectively, as typical values slightly deviating from $\Delta=1$ of the isotropic Heisenberg case. From the MC simulations (see Appendix), the transition temperature in each case is estimated to be $T_N/|J| \simeq 0.75$ for $\Delta=1.05$ and $T_{KT}/|J| \simeq 0.6 $ for $\Delta=0.95$ \cite{KT_XXZ_Cuccoli_95, KT_XXZ_Lee_05, KT_XXZ_Pires_96}.

In Eq. (\ref{eq:conductivity}), the temperature dependence of $\kappa_{\mu\nu}$ is determined by the integrated value of the time correlation of the thermal current $\langle J_{th,\nu}(0) \, J_{th,\mu}(t) \rangle$ except the trivial $T^{-2}$ factor, so that we will start from the temperature dependence of $\langle J_{th,\nu}(0) \, J_{th,\mu}(t) \rangle$. Figure \ref{fig:timedep_thermal_comp} shows the time correlation function normalized by the system size $\langle j_{th,x}(0) \, j_{th,x}(t) \rangle \equiv \langle J_{th,x}(0) \, J_{th,x}(t) \rangle/L^2$ at different temperatures in the Ising-type ($\Delta = 1.05$), $XY$-type ($\Delta = 0.95$), and Heisenberg-type ($\Delta = 1$) spin systems. System-size dependence can hardly be seen, suggesting that the thermal transport is a spatially local phenomenon. As for the effect of the magnetic anisotropy, there is no qualitative difference among the three cases. With decreasing temperature, the time correlation decays more slowly in time. In other words, the relaxation time of the thermal current, which we denote as $\tau_{th}$, becomes longer. Thus, the associated thermal conductivity $\kappa_{\mu \nu}$ is expected to follow a common monotonic temperature-dependence.      

Figure \ref{fig:thermal_comp} shows the longitudinal and transverse thermal conductivities as a function of temperature $T$ in the Ising-type ($\Delta = 1.05$), $XY$-type ($\Delta = 0.95$), and Heisenberg-type ($\Delta = 1$) spin systems. Because the $yy$ ($xy$) component of $\kappa_{\mu \nu}$ is equivalent to the $xx$ ($yx$) component in the present square-lattice NN model, only the the $xx$ and $yx$ components, $\kappa_{xx}$ and $\kappa_{yx}$, are shown. One can see from Fig. \ref{fig:thermal_comp} that in all the three cases, the transverse Hall response $\kappa_{yx}$ is absent at $2\sigma$ precision (see lower panels) and the longitudinal thermal conductivity $\kappa_{xx}$ gradually increases toward $T=0$ (see the upper main panels). Although the phase transition occurs in the anisotropic spin systems, no clear anomaly can be seen in the thermal conductivity at the magnetic transition temperature $T_N$ or the KT topological transition temperature $T_{KT}$. Thus, in view of the main focus of this work, our conclusion is that the strong association between the thermal conductivity and the phase transition cannot be observed in the present NN XXZ model in two dimensions.
Below in this section, to shed light on the basic properties of the thermal transport in the classical spin systems, we will devote ourselves to the low-temperature behavior of the longitudinal thermal conductivity $\kappa_{xx}$.

\begin{figure}[t]
\includegraphics[scale=1.0]{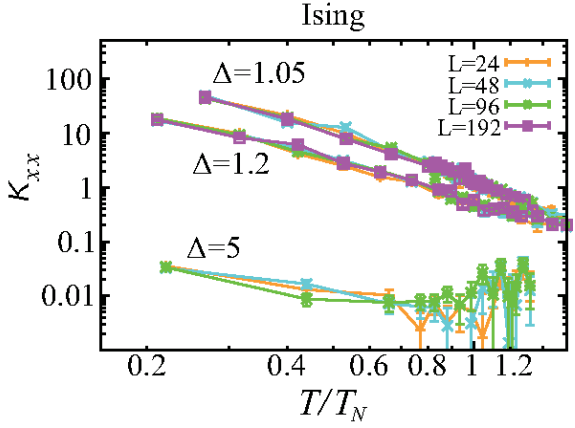}
\caption {The log-log plot of the longitudinal thermal conductivity $\kappa_{x x}$ as a function of $T/T_{N}$ in the cases of the Ising anisotropies of $\Delta = 1.05$ (top), $\Delta = 1.2$ (middle), and $\Delta = 5$ (bottom). \label{fig:thermal_Ising}}
\end{figure}

For the $XY$-type anisotropy $\Delta<1$, the temperature dependence of $\kappa_{xx}$ in Fig. \ref{fig:thermal_comp} (b) is not altered qualitatively by the change in $\Delta$. For the Ising-type anisotropy $\Delta >1$, on the other hand, the magnon excitation has the gap $\Delta_{gp}=4|J|S \sqrt{\Delta^2-1}$, so that the thermal current, which is the energy flow carried by the manons, and the associated conductivity $\kappa_{xx}$ are expected to be suppressed with increasing $\Delta$. Figure \ref{fig:thermal_Ising} shows the longitudinal thermal conductivity $\kappa_{xx}$ as a function of $T/T_N$ for various values of $\Delta > 1$. Not only the absolute value of $\kappa_{xx}$ but also the divergent behavior toward $T=0$ is suppressed by the increase of $\Delta$. At least for not so strong Ising-type anisotropy, however, $\kappa_{xx}$ tends to diverge toward $T=0$, roughly showing a power-law behavior. Hereafter, we will discuss the origin of such a power-law-type temperature dependence, focusing on the almost isotropic spin systems. 

As one can see from Eq. (\ref{eq:conductivity}), $\kappa_{xx}$ involves the trivial $T^{-2}$ dependence. In order to extract the nontrivial temperature dependence other than the $T^{-2}$ factor, $T^2 \kappa_{xx}=\int dt \, \langle j_{th,x}(0) \, j_{th,x}(t) \rangle$ is plotted in the insets of the upper panels of Fig. \ref{fig:thermal_comp} as a function of temperature. In the anisotropic cases of $\Delta \neq 1$, $T^2 \kappa_{xx}$ tends to saturate to a constant value at the lowest temperature, whereas in the isotropic case of $\Delta=1$, it remains increasing toward $T=0$. Except this difference at the lowest temperature, $T^2 \kappa_{xx}$ shows a weak monotonic increase below $T/|J|\leq 0.8$ in both the anisotropic and isotropic cases. Thus, the divergent behavior toward $T=0$ in $\kappa_{xx}$ is mainly due to the $T^{-2}$ factor, but in the low-temperature range of our simulations, $\kappa_{xx}$ increases slightly faster than $T^{-2}$ due to the non-trivial contribution originating from the thermal fluctuation, $T^2 \kappa_{xx}=\int dt \, \langle j_{th,x}(0) \, j_{th,x}(t) \rangle$.  
The analytical result in Eq. (\ref{eq:conductivity_classical_th}), on the other hand, shows that the thermal conductivity due to the magnon propagation should behave as $\kappa_{xx} \propto 1/\alpha \propto T^{-2}$. As mentioned above, at least in the temperature range of our simulations, the numerically obtained $\kappa_{xx}$ increases faster than $T^{-2}$. In order to examine the origin of the deviation between the numerical and analytical results on the temperature dependence of $\kappa_{xx}$, we will look into the details of the temperature dependences of the physical quantities related to $\langle j_{th,x}(0) \, j_{th,x}(t) \rangle$. 


\begin{figure}[t]
\includegraphics[scale=0.9]{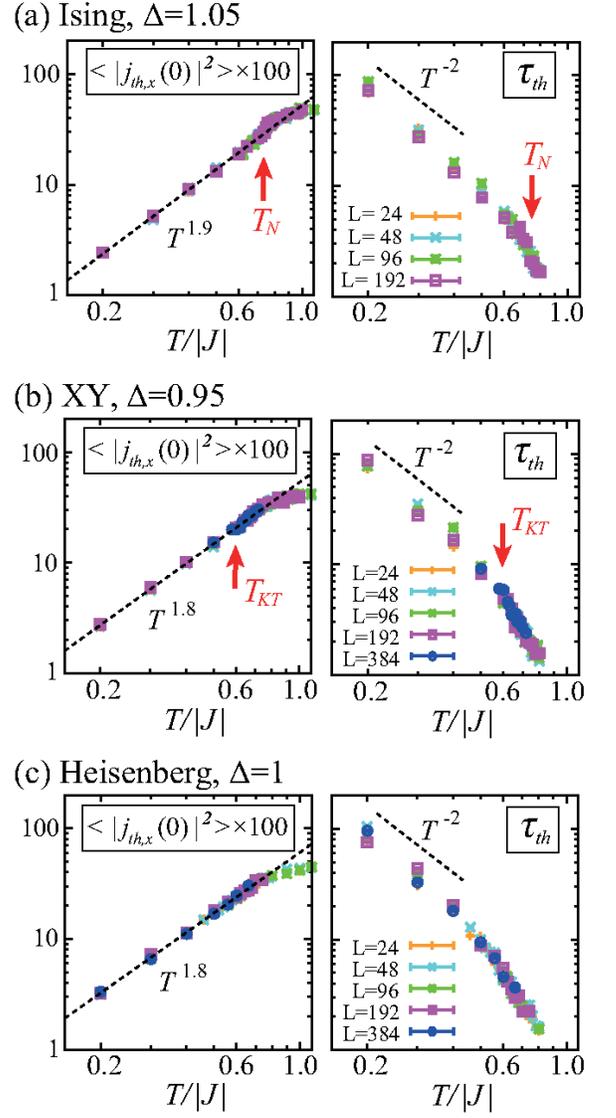}
\caption {The log-log plots of the temperature dependences of $\langle | j_{th,x}(0)|^2 \rangle$ (left panels) and the relaxation time of the thermal current $\tau_{th}$ (right panels) in the (a) Ising-type ($\Delta=1.05$), (b) $XY$-type ($\Delta=0.95$), and (c) Heisenberg-type ($\Delta=1$) spin systems. $\tau_{th}$ and $\langle |j_{th,x}(0)|^2 \rangle$ are measured in units of $|J|^{-1}$ and $|J|^4$, respectively. In the left panels, $\langle | j_{th,x}(0)|^2 \rangle$ is multiplied by $100$ such that the scale of the vertical axis be the same as that in the right panels. In the left panels, a dashed curve represents a power function of $T$ obtained by fitting the low-temperature data in each case, and in the right panels, the analytically expected $T^{-2}$ dependence is presented for reference. \label{fig:thermal_tau_comp}}
\end{figure}

In Fig. \ref{fig:timedep_thermal_comp}, the time correlation $\langle j_{th,x}(0) \, j_{th,x}(t) \rangle$ decays exponentially in the form of $e^{- t/\tau_{th}}$ with the relaxation time of the thermal current $\tau_{th}$, so that we could assume $\langle j_{th,x}(0) \, j_{th,x}(t) \rangle \simeq \langle |j_{th,x}(0)|^2\rangle e^{-t/\tau_{th}}$. Then, by carrying out the integral over time in Eq. (\ref{eq:conductivity}), one can estimate the longitudinal thermal conductivity as $\kappa_{xx} \simeq T^{-2} \, \langle | j_{th,x}(0)|^2 \rangle \, \tau_{th}$. As the data on the static quantity $\langle | j_{th,x}(0)|^2 \rangle$ can be compared directly with the analytical result given in Eq. (\ref{eq:Jth_static_final}), one can relate $\tau_{th}$ to the magnon damping $\alpha$ via Eq. (\ref{eq:conductivity_classical_th}). If the equal-time correlation $\langle | j_{th,x}(0)|^2 \rangle$ follows the $T^2$ dependence expected in LSWT, the relaxation time of the thermal current $\tau_{th}$ corresponds to the inverse magnon-damping $1/\alpha$ which is roughly proportional to $T^{-2}$ in the lowest-order approximation \cite{MagnonDamping_Tyc_89, MagnonDamping_Harris_71}. 

Figure \ref{fig:thermal_tau_comp} shows the temperature dependences of $\langle | j_{th,x}(0)|^2 \rangle$ and $\tau_{th}$ in the three cases of $\Delta=1.05$, $\Delta=0.95$, and $\Delta=1$, where $\tau_{th}$ is extracted by fitting the $\langle j_{th,x}(0) \, j_{th,x}(t) \rangle$ curve with the exponential form of $e^{-t/\tau_{th}}$. Since $\langle | j_{th,x}(0)|^2 \rangle$ exhibits a power-law behavior, we fit the low-temperature data with the functional form of $T^x$ and find $x=1.8\sim 1.9$. The resultant fitting function $T^x$ in each case is represented by a dashed curve together with the obtained value of $x$ in Fig. \ref{fig:thermal_tau_comp}. The exponent $x\simeq 2$ for $\langle | j_{th,x}(0)|^2 \rangle$ is in good agreement with the analytical result given in Eq. (\ref{eq:Jth_static_final}), so that the origin of the discrepancy in the temperature dependence of $\kappa_{xx}$ between the numerical and analytical results consists in the relaxation time $\tau_{th}$ which should satisfy the relation $\tau_{th} \propto 1/\alpha \propto T^{-2}$. As one can see from the right panels in Fig. \ref{fig:thermal_tau_comp}, however, $\tau_{th}$ diverges toward $T=0$ slightly faster than $T^{-2}$. A rough estimation, which is done by fitting all the low-temperature data for $T/|J| \leq 0.6$ with the functional form of $T^x$, yields $\tau_{th} \propto T^{-2.5}$ in all the three cases. The deviation from the expected behavior $1/\alpha \propto T^{-2}$ may be attributed to the temperature range considered. The temperature range available for fitting might be higher than that assumed in the analytical calculation where higher-order multi-magnon-scattering processes are neglected. With further decreasing temperature below the lowest temperature of our simulation, $\tau_{th}$ and resultant $\kappa_{xx}$ should tend to obey the expected power-law form $T^{-2}$. Actually, in the Ising and XY cases, a precursor of such a tendency has already been observed as the saturated behavior in $T^2 \kappa_{xx}$ (see the insets of Fig. \ref{fig:thermal_comp}).
  

\section{Numerical results on the spin-current conductivity}
\begin{figure}[t]
\includegraphics[width=\columnwidth]{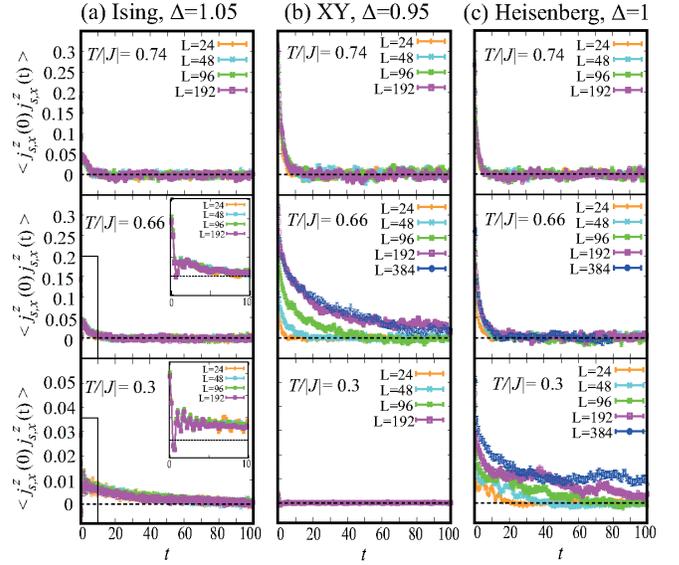}
\caption{The time correlation function of the spin current $\langle j^z_{s,x}(0) \, j^z_{s,x}(t) \rangle$ at $T/|J|=0.74$ (top), $T/|J|=0.66$ (middle), and $T/|J|=0.3$ (bottom) in the (a) Ising-type ($\Delta=1.05$), (b) $XY$-type ($\Delta=0.95$), and (c) Heisenberg-type ($\Delta=1$) spin systems. Time $t$ and $\langle j^z_{s,x}(0) \, j^z_{s,x}(t) \rangle$ are measured in units of $|J|^{-1}$ and $|J|^2$, respectively. In (a), the inset shows the zoomed view of the short-time region near $t=0$ enclosed by a box in each main panel. \label{fig:timedep_spin_comp}}
\end{figure}

\begin{figure*}[t]
\includegraphics[scale=0.78]{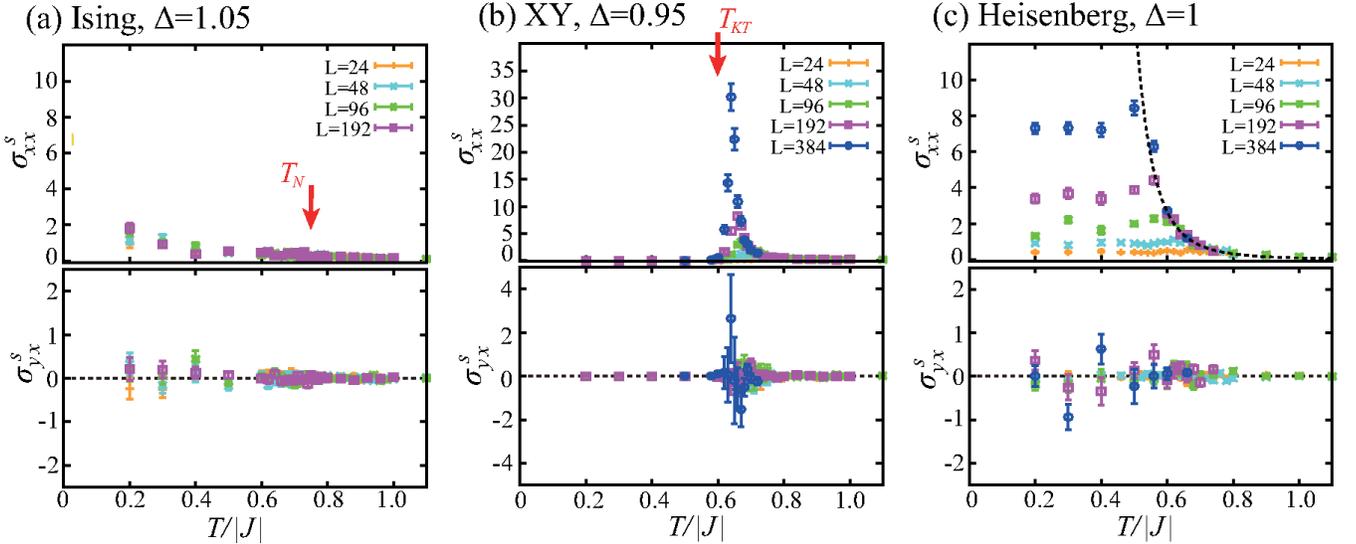}
\caption{The temperature dependence of the spin-current conductivity $\sigma^s_{\mu \nu}$ in the (a) Ising-type ($\Delta=1.05$), (b) $XY$-type ($\Delta=0.95$), and (c) Heisenberg-type ($\Delta=1$) spin systems, where upper and lower panels show the longitudinal and transverse conductivities, respectively. In (a) and (b), red arrows indicate the magnetic and KT transition temperatures, $T_N/|J|\simeq 0.75$ and $T_{KT}/|J| \simeq 0.6$, respectively. In (c), a dashed curve in the upper panel represents the $\sigma^s_{xx}(T)$ curve extrapolated to the thermodynamic limit of $L \rightarrow \infty$ (see the main text). \label{fig:spin_comp}}
\end{figure*}

In Sec. III, based on the analytical calculations in LSWT, we find that the effect of the magnetic anisotropy $\Delta$, i.e., the difference in the ordering properties, is reflected in the low-temperature spin-transport. In this section, we will discuss the association between the phase transition and the spin-current conductivity $\sigma^s_{\mu \nu}$, based on numerical results. 

We shall start from the time correlation function of the spin current $\langle J^z_{s,\nu}(0) \, J^z_{s,\mu}(t) \rangle$ which yields the nontrivial temperature dependence of $\sigma^s_{\mu \nu}$ [see Eq. (\ref{eq:conductivity})]. Figure \ref{fig:timedep_spin_comp} shows the time correlation function normalized by the system size $\langle j^z_{s,x}(0) \, j^z_{s,x}(t) \rangle=\langle J^z_{s,x}(0) \, J^z_{s,x}(t) \rangle/L^2$ at various temperatures in the typical three cases, Ising-type ($\Delta = 1.05$), $XY$-type ($\Delta = 0.95$), and Heisenberg-type ($\Delta=1$) spin systems. These $\Delta$ values are the same as those in Figs. \ref{fig:timedep_thermal_comp} and \ref{fig:thermal_comp}. 
At the high temperature $T/|J|=0.74$, one cannot see a clear difference among the three cases. With decreasing temperature, $\langle j^z_{s,x}(0) \, j^z_{s,x}(t) \rangle$ exhibits characteristic behaviors depending on the ordering properties. In the Ising case of $\Delta=1.05$, $\langle j^z_{s,x}(0) \, j^z_{s,x}(t) \rangle$ shows an oscillating behavior in the very-short time scale [see the insets of Fig. \ref{fig:timedep_spin_comp}(a)], but its long-time relaxation whose characteristic time scale is denoted by $\tau_s$ becomes slower at lower temperatures without showing the system size dependence. In the $XY$ case of $\Delta=0.95$, the time correlation persists for a long time at $T/|J|=0.66$ slightly above $T_{KT}$, showing a large system size dependence, whereas the time correlation is lost within a short time scale at $T/|J|=0.3$ much lower than $T_{KT}$. In the Heisenberg case of $\Delta=1$, $\tau_s$ becomes longer with decreasing temperature like in the Ising case, but the system size dependence is quite large. The above difference is reflected in the spin-current conductivity $\sigma^s_{\mu \nu}$ through the integration of $\langle j^z_{s,\nu}(0) \, j^z_{s,\mu}(t) \rangle$ over the whole time range.

Figure \ref{fig:spin_comp} shows the temperature dependences of the longitudinal (upper panels) and transverse (lower panels) spin-current conductivities, $\sigma^s_{xx}$ and $\sigma^s_{yx}$, for $\Delta=1.05$ (a), $\Delta=1.05$ (b), and $\Delta=1$ (c). As one can see from Fig. \ref{fig:spin_comp}, in all the three cases, the transverse Hall response $\sigma^s_{yx} \, (=\sigma^s_{xy})$ is absent also for the spin transport as well as the thermal transport. The longitudinal spin-current conductivity $\sigma^s_{xx} \, (=\sigma^s_{yy})$, on the other hand, exhibits temperature dependences characteristic of the three different universality classes. 
Here, we briefly summarize the temperature dependence of $\sigma^s_{xx}$, and a detailed analysis in each case will be given in the following subsections. In the Ising case of $\Delta=1.05$, $\sigma^s_{xx}$ gradually increases with decreasing temperature without showing a clear anomaly at the magnetic transition temperature $T_N$. Also, the system size dependence cannot be seen, as is already suggested from the size-independent time-correlation-functions in Fig. \ref{fig:timedep_spin_comp} (a). In the $XY$ case of $\Delta=0.95$, $\sigma^s_{xx}$ exhibits a divergent sharp peak toward the KT transition temperature $T_{KT}$, and becomes vanishingly small at lower temperatures below $T_{KT}$. In the Heisenberg case of $\Delta=1$, $\sigma^s_{xx}$ increases exponentially with decreasing temperature, showing a large system-size-dependence at lower temperatures. 
Below in this section, we will give a detailed description of the association between the longitudinal spin-current conductivity $\sigma^s_{xx}$ and the ordering properties of the system.

\subsection{Ising-type spin system}
In Fig. \ref{fig:spin_comp}, for the Ising-type anisotropy of $\Delta = 1.05$, a clear signature of the magnetic transition at $T_N$ cannot be seen in $\sigma^s_{xx}$. We will first check that this result is not altered qualitatively by the value of $\Delta$, and subsequently discuss the temperature dependence of $\sigma^s_{xx}$ in the long-range-ordered phase below $T_N$, making a comparison between the numerical result and the analytical one in Sec. III.

\begin{figure}[t]
\includegraphics[scale=0.95]{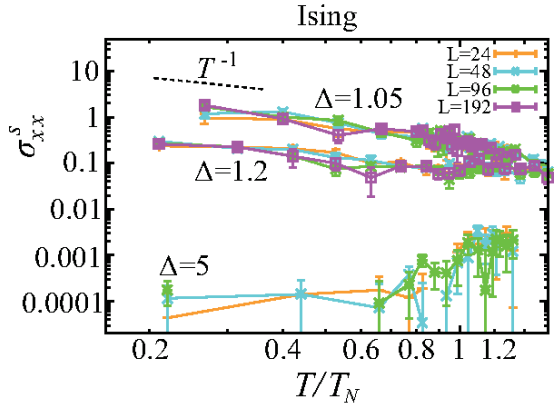}
\caption {The log-log plot of the longitudinal spin-current conductivity $\sigma^s_{x x}$ as a function of $T/T_{N}$ in the cases of the Ising anisotropies of $\Delta = 1.05$ (top), $\Delta = 1.2$ (middle), and $\Delta = 5$ (bottom). The $T^{-1}$ dependence expected for an almost isotropic case is presented for reference (see the main text).\label{fig:spin_Ising}}
\end{figure}

The gap-opening in the magnon excitation due to $\Delta$ is expected to suppress $\sigma^s_{xx}$, as is actually the case for the thermal conductivity $\kappa_{xx}$. Figure \ref{fig:spin_Ising} shows $\sigma^s_{xx}$ as a function of $T/T_N$ for various values of $\Delta > 1$. No clear signature of the magnetic transition can commonly be seen near $T_N$, and as is expected, $\sigma^s_{xx}$ is suppressed by the increase of $\Delta$. For relatively weak magnetic anisotropies, $\sigma^s_{xx}$ increases toward $T=0$ and its temperature dependence is almost compatible with the analytical expectation, $\sigma^s_{xx} \propto T/\alpha \propto T^{-1}$, given in Eq. (\ref{eq:conductivity_classical_spin}). To look into the details of the temperature effect on $\sigma^s_{xx}$, we will examine the temperature dependence of the current-related quantities for $\Delta=1.05$.

In Fig. \ref{fig:timedep_spin_comp} (a), except for the short-time oscillating behavior, the time correlation $\langle j^z_{s,x}(0) \, j^z_{s,x}(t) \rangle$ decays exponentially in the form of $e^{-t/\tau_s }$, so that we could roughly write $\langle j^z_{s,x}(0) \, j^z_{s,x}(t) \rangle \sim \langle |j^z_{s,x}(0)|^2 \rangle \, e^{-t/\tau_s }$. Then, from Eq. (\ref{eq:conductivity}), the longitudinal spin-current conductivity $\sigma^s_{xx}$ can be evaluated as $\sigma^s_{xx} \sim T^{-1} \, \langle |j^z_{s,x}(0)|^2 \rangle \, \tau_s $. If the static quantity $\langle |j^z_{s,x}(0)|^2 \rangle$ follows the $T^2$ behavior expected in LSWT [see Eq. (\ref{eq:Js_static})] as is actually the case for the thermal transport, it follows that $\sigma^s_{xx} \sim T \tau_s$. By comparing this expression to Eq. (\ref{eq:conductivity_classical_spin}), one notice that $\tau_s$ is associated with the magnon damping $\alpha$ via $\tau_s \sim 1/\alpha$. 

\begin{figure}[t]
\includegraphics[scale=0.9]{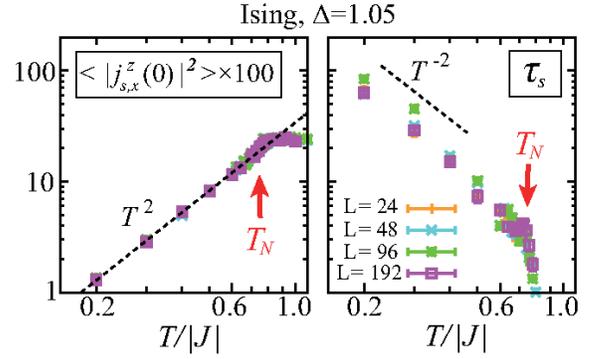}
\caption {The temperature dependences of the equal-time spin-current correlation $\langle | j^z_{s,x}(0)|^2 \rangle$ (left panel) and the relaxation time of the spin current $\tau_s$ (right panel) in the case of the Ising anisotropy of $\Delta = 1.05$, where a red arrow indicates the magnetic transition temperature $T_N$. In the left panel, a dashed curve represents a power function of $T$ obtained by fitting the low-temperature data, and in the right panel, the analytically expected $T^{-2}$ dependence is presented for reference. $\tau_{s}$ and $\langle |j^z_{s,x}(0)|^2 \rangle$ are measured in units of $|J|^{-1}$ and $|J|^2$, respectively. In the left panel, $\langle | j^z_{s,x}(0)|^2 \rangle$ is multiplied by $100$ such that the scale of the vertical axis be the same as that in the right panel.\label{fig:spin_tau_Ising}}
\end{figure}

Figure \ref{fig:spin_tau_Ising} shows the temperature dependences of $\langle | j^z_{s,x}(0)|^2 \rangle$ and $\tau_s$, where $\tau_s$ is extracted by fitting the tail of $\langle j^z_{s,x}(0) \, j^z_{s,x}(t) \rangle$ with $e^{-t/\tau_s}$. As one can see from the left panel of Fig. \ref{fig:spin_tau_Ising}, $\langle | j^z_{s,x}(0)|^2 \rangle$ shows a power-law behavior of the form $T^x$ in the ordered phase, and the exponent $x$ is obtained by fitting the low-temperature data as $x=2$. The resultant fitting function is represented by a dashed curve in the left panel of Fig. \ref{fig:spin_tau_Ising}. The obtained $T^2$ behavior for $\langle | j^z_{s,x}(0)|^2 \rangle$ is in good agreement with the analytical result in Eq. (\ref{eq:Js_static}), so that $\tau_s \sim 1/\alpha \propto T^{-2}$ should be satisfied. The numerically obtained $\tau_s$ shown in the right panel of Fig. \ref{fig:spin_tau_Ising} tends to obey the expected power-law form $T^{-2}$, but in the wide low-temperature range of our simulation, it increases toward $T=0$ slightly faster than $T^{-2}$. When we fit all the low-temperature data below $T/|J|=0.6$ with the functional form $T^x$, the same temperature dependence as that of the thermal-current-relaxation time $\tau_{th}$ is obtained for the spin-current-relaxation time, namely, $\tau_s \propto T^{-2.5}$, indicating that in the Ising-type spin systems, the long-time relaxations of the spin and thermal transports are of the same origin, namely, the magnon damping due to the multi-magnon scatterings. 

In the short-time scale, on the other hand, one can see the oscillating behavior in $\langle j^z_{s,x}(0) \, j^z_{s,x}(t) \rangle$ [see the insets in Fig. \ref{fig:timedep_spin_comp} (a)], which is not observed in the thermal-current relaxation. Although the origin of the oscillation is not clear, this suggests that the spin-current relaxation may involve not only the ordinary magnon damping but also other effects of the magnetic excitations. As we will see below, in the $XY$-type spin systems, the vortex excitations come into play in the spin-current relaxation, leading to the divergence of $\sigma^s_{xx}$ at the KT transition temperature.


\subsection{$XY$-type spin system}
\begin{figure}[t]
\includegraphics[scale=1.0]{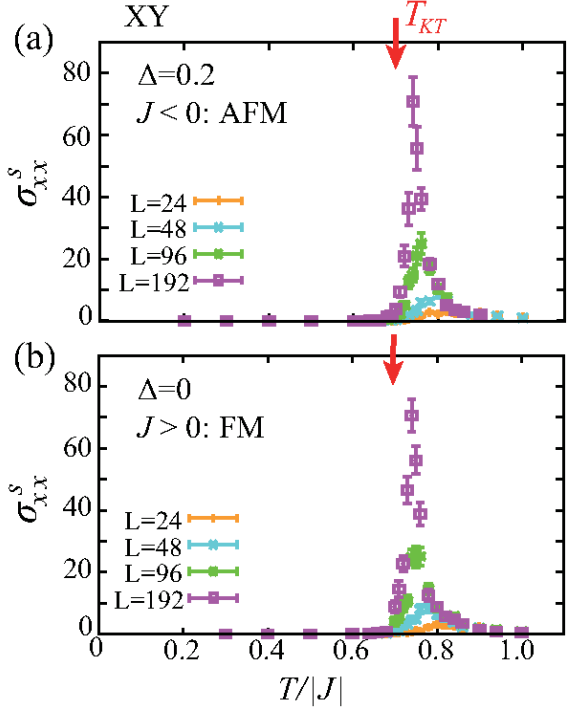}
\caption {The temperature dependence of the longitudinal spin-current conductivity $\sigma^s_{x x}$ in the $XY$-type antiferromangnet ($J<0$) with $\Delta = 0.2$ (a), and ferromagnet ($J>0$) with $\Delta = 0$ (b). A red arrow indicates the KT transition temperature in the thermodynamic limit, $T_{KT}/|J| \simeq 0.7$ \cite{KT_XXZ_Cuccoli_95, KT_XXZ_Lee_05, KT_XXZ_Pires_96}. \label{fig:spin_XY}}
\end{figure}

In the $XY$ antiferromagnet with the weak anisotropy $\Delta =0.95$, as shown in Fig. \ref{fig:spin_comp} (b), the longitudinal spin-current conductivity $\sigma^s_{xx}$ (=$\sigma^s_{yy}$) is significantly enhanced near $T_{KT}$, but once entering in the low-temperature phase below $T_{KT}$, $\sigma^s_{xx}$ becomes vanishingly small. These features are universal in the $XY$-type spin systems, being independent of the values of $\Delta$. Furthermore, even if the antiferromagnetic exchange interaction $J <0$ is replaced with a ferromagnetic one $J>0$, the universality class remains unchanged and the above features in $\sigma^s_{xx}$ can be observed. Figure \ref{fig:spin_XY} shows the temperature dependence of $\sigma^s_{xx}$ in the antiferromagnet ($J<0$) with $\Delta=0.2$ (a) and in the ferromagnet ($J>0$) with $\Delta=0$ (b). In both cases, a divergent sharp peak can clearly be seen near $T_{KT}$. With increasing the system size $L$, the peak height increases and the peak temperature approaches $T_{KT}$ from above, suggesting that in the thermodynamic limit of $L \rightarrow \infty$, $\sigma^s_{xx}$ diverges at $T_{KT}$. On crossing $T_{KT}$ from above, $\sigma^s_{xx}$ drops to a vanishingly small value. Hereafter, we will discuss the origin of this temperature dependence.

\begin{figure}[t]
\includegraphics[scale=1.0]{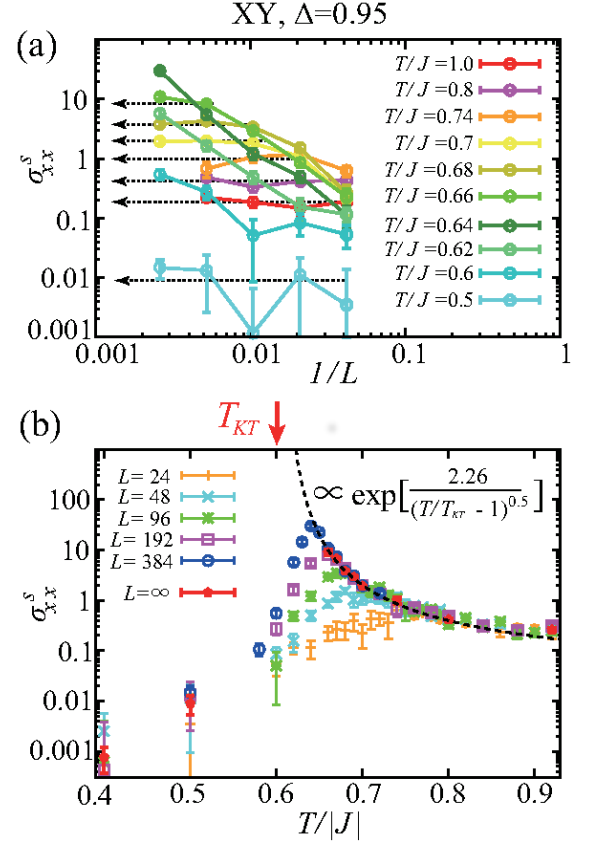}
\caption {The longitudinal spin-current conductivity $\sigma^s_{x x}$ in the $XY$ case of $\Delta = 0.95$. (a) The system-size dependence of $\sigma^s_{x x}$ at various temperatures and (b) the semi-logarithmic plot of the temperature dependence of $\sigma^s_{x x}$. In (a), an arrow represents the extrapolated $L \rightarrow \infty$ value at each temperature. In (b), a red arrow indicates $T_{KT}$ and a dashed curve represents the $\sigma^s_{xx}(T)$ curve extrapolated to the thermodynamic limit (see the main text). \label{fig:spin_Linf_XY}}
\end{figure}

As discussed in Sec. III, in the ordered phase of the $XY$-type spin system, the leading-order magnon-spin-current is absent [see Eq. (\ref{eq:current_spin_mag})] because of the orthogonal relation between the quantization axis lying in the $xy$-plane of the spin space and the polarization direction of the spin current which is in the $z$ direction in the present XXZ model. The associated spin-current conductivity $\sigma^s_{xx}$, therefore, should be vanishingly small, although higher-order magnon contributions may have a little effect on the spin transport. The low-temperature feature observed below $T_{KT}$ in Figs. \ref{fig:spin_comp} (b) and \ref{fig:spin_XY} is understood as a manifestation of this nature inherent to the $XY$-type anisotropy. Thus, the non-trivial issue is the significant enhancement of $\sigma^s_{xx}$ near $T_{KT}$ observed in the numerical simulations.  

Since the system-size-dependent divergent peak near $T_{KT}$ is commonly observed for the $XY$-type anisotropy, we focus on the case of $\Delta=0.95$ as a representative example and discuss the thermodynamic limit ($L \rightarrow \infty$) of $\sigma^s_{xx}$. Figure \ref{fig:spin_Linf_XY} (a) shows the system-size dependence of $\sigma^s_{xx}$ at various temperatures. 
One can see that at temperatures away from $T_{KT}/|J|\simeq 0.6$, $\sigma^s_{xx}$ as a function of the system size $L$ saturates to a constant value, which corresponds to $\sigma^s_{xx}$ in the thermodynamic limit.
The extrapolated $L \rightarrow \infty$ value of $\sigma^s_{xx}$ and the corresponding original finite-size data in Fig. \ref{fig:spin_comp} (b) are plotted in Fig. \ref{fig:spin_Linf_XY} (b) on the semi-logarithmic scale. The divergent behavior toward $T_{KT}/|J| \simeq 0.6$ and the sudden drop across $T_{KT}$ can clearly be seen. Noting that the spin correlation length $\xi_s$ in the KT transition is known to diverge in the form of $\xi_s/a \sim \exp\big[b_{KT} /\sqrt{T/T_{KT}-1} \big]$ with $b_{KT} \simeq \pi/2$ \cite{KT_Kosterlitz_74}, we fit the $L \rightarrow \infty$ data of $\sigma^s_{xx}$ at $T \gtrsim T_{KT}$ with the functional form of $A \exp\big[ B/\sqrt{T/T_{KT}-1} \big]$. The fitting parameters $A$ and $B$ are obtained as $B=2.26 \pm 0.10$ and $A=0.008 \pm 0.002$. The $\sigma^s_{xx}(T)$ curve extrapolated in this way is represented by a dashed curve in Fig. \ref{fig:spin_Linf_XY} (b). One can see that the obtained exponential form well characterizes the numerically-obtained divergent behavior of $\sigma^s_{xx}$, which, together with the obtained $B$-value comparable to $b_{KT}$, suggests that this pronounced spin-transport phenomenon is closely related to the KT transition, or equivalently, the vortex binding-unbinding process. 

\begin{figure}[t]
\includegraphics[scale=1.0]{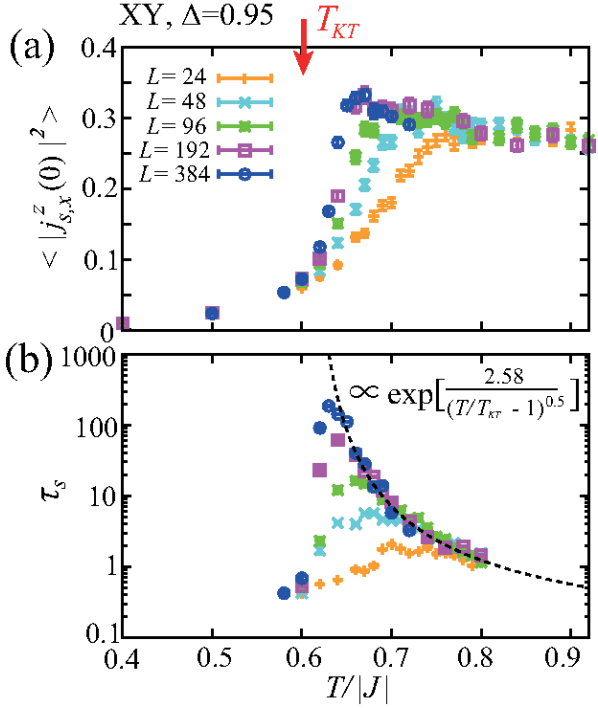}
\caption {The temperature dependences of the equal-time spin-current correlation $\langle | j^z_{s,x}(0)|^2 \rangle$ (a), and the spin-current relaxation time $\tau_s$ (b), in the case of the $XY$-type anisotropy of $\Delta = 0.95$, where these quantities are measured in the same units as those in Fig. \ref{fig:spin_tau_Ising}. A red arrow indicates $T_{KT}$ and a dashed curve in (b) represents an exponential function obtained by fitting the data above $T_{KT}$. \label{fig:spin_tau_XY}}
\end{figure}

In the KT transition, the spin correlation length $\xi_s$ corresponds to the inter-free-vortex distance. With decreasing temperature above $T_{KT}$, the inter-free-vortex distance increases, so that it becomes difficult for a single vortex to find out a partner free anti-vortex to form a vortex pair. This means that in terms of the time evolution, the single free vortex wanders for a longer time until it collides with the partner free anti-vortex. Thus, the lifetime of the single free vortex should get longer on approaching $T_{KT}$ from above. Once across $T_{KT}$, all the vortices are paired up and a single vortex cannot be found any more. Bearing this fundamental physics of the KT transition in our mind, we examine the temperature dependences of $\tau_s$ and $\langle | j^z_{s,x}(0)|^2 \rangle$. 

Figure \ref{fig:spin_tau_XY} shows the temperature dependences of $\langle | j^z_{s,x}(0)|^2 \rangle$ (a) and $\tau_s$ (b). The spin-current relaxation time $\tau_s$ is determined by fitting the long-time tail of $\langle j^z_{s,x}(0)j^z_{s,x}(t) \rangle$ in Fig. \ref{fig:spin_Ising} (b) with the exponential form of $e^{-t/\tau_s}$. One can see from Fig. \ref{fig:spin_tau_XY} that on approaching $T_{KT}$ from above, $\tau_s$ is significantly enhanced, while $\langle | j^z_{s,x}(0)|^2 \rangle$ only shows a slight increase. In the low-temperature phase below $T_{KT}$, $\langle | j^z_{s,x}(0)|^2 \rangle$ is strongly suppressed as is expected from the analytical result that the leading-order magnon-spin-current is absent, and correspondingly, the relaxation becomes so rapid that $\tau_s$ cannot be defined any more. The functional type characterizing the steep increase in $\tau_s$ is also the exponential one. By fitting the data at $T \gtrsim T_{KT}$ with the form of $\tilde{A} \exp\big[ \tilde{B}/\sqrt{T/T_{KT}-1} \big]$, we obtain $\tilde{A}=0.013 \pm 0.003$ and $\tilde{B}=2.58 \pm 0.07$. The extrapolated $\tau_s(T)$ curve is represented by a dashed curve in Fig. \ref{fig:spin_tau_XY}. One can see that the obtained exponential form well characterizes the numerically-obtained divergent behavior of $\tau_s$. As $\sigma^s_{xx}$ is related to $\tau_s$ and $\langle | j^z_{s,x}(0)|^2 \rangle$ via $\sigma^s_{xx}\sim T^{-1} \, \langle | j^z_{s,x}(0)|^2 \rangle \, \tau_s$, the divergent behavior in $\sigma^s_{xx}$ originates from the divergence of the spin-current-relaxation time $\tau_s$ toward $T_{KT}$. Actually, the obtained values of $B\simeq 2.26$ and $\tilde{B}\simeq 2.58$ almost coincide with each other.

Now, we will address the physical interpretation of the above result. 
In the KT topological transition, the distinct feature above $T_{KT}$ is the existence of an isolated free vortex and its dynamics. Since the vortex interacts with surrounding magnons or spin waves, the vortex motion is not ballistic, but rather diffusive \cite{KT-diffusive_Loft_87,KT-diffusive_Toyoki_90,KT-diffusive_Goldenfeld_90,KT-diffusive_Huse_93,KT-diffusive_Bray_00}. Thus, the vortex lifetime $\tau_{vtx}$ could be estimated roughly as $\tau_{vtx} \propto \xi_s^2 \sim \exp\big[ 2b_{KT} /\sqrt{T/T_{KT}-1} \big]$, so that $\tau_{vtx}$ should get longer in the exponential form toward $T_{KT}$ with $2b_{XT} \simeq \pi$.     
Since the two time-scales, $\tau_s$ and $\tau_{vtx}$, develop toward $T_{KT}$ in the almost same manner as a function of temperature, it is naturally expected that the vortex excitations play an important role in the spin-current relaxation. 
Because $\sigma^s_{xx}$ is proportional to $\tau_s$, we could conclude that the divergent peak at $T_{KT}$ in the $\sigma^s_{xx}$ curve is attributed to the topological excitations of the long-life-time vortices.  

\subsection{Heisenberg-type spin system}
\begin{figure}[t]
\includegraphics[scale=1.0]{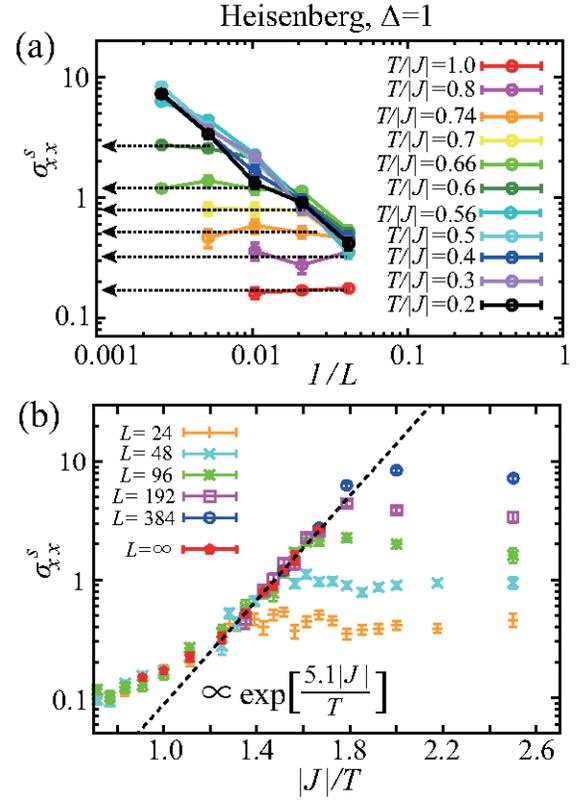}
\caption {The longitudinal spin-current conductivity $\sigma^s_{x x}$ in the Heisenberg case of $\Delta = 1$. (a) The system-size dependence of $\sigma^s_{x x}$ at various temperatures and (b) the semi-log plot of $\sigma^s_{x x}$ as a function of $1/T$. In (a), an arrow represents the extrapolated $L \rightarrow \infty$ value at each temperature. In (b), a dashed curve represents the $\sigma^s_{xx}(T)$ curve extrapolated to the thermodynamic limit (see the main text). \label{fig:spin_Linf_Heisenberg}}
\end{figure}

In the Heisenberg case of $\Delta=1$, the spin space is isotropic, so that in contrast to the anisotropic cases of $\Delta \neq 1$, not only the $z$ component of the magnetization but also the $x$ and $y$ components are conserved quantities. This enables one to define the spin-currents ${\bf J}_s^x$ and ${\bf J}_s^y$ as well as ${\bf J}_s^z$, where ${\bf J}_s^\alpha = J\sum_{\langle i,j \rangle} \big({\bf r}_i-{\bf r}_j \big) \big( {\bf S}_i \times {\bf S}_j \big)^\alpha$ can be derived in the same manner as Eq. (\ref{eq:j_sc}). Since all the spin currents, ${\bf J}_s^x$, ${\bf J}_s^y$, and ${\bf J}_s^z$, should be equivalent to one another, the associated spin-current conductivities should also be equivalent. Thus, in the Heisenberg case, we calculate the spin-current conductivity averaged over the three spin components
\begin{eqnarray}\label{eq:conductivity_Heisenberg}
\sigma_{\mu \nu}^s &=& \frac{1}{T \, L^2} \int_0^\infty dt \, \frac{1}{3} \Big( \big\langle J^x_{s,\nu}(0) \, J^x_{s,\mu}(t) \big\rangle \nonumber\\
&& + \big\langle J^y_{s,\nu}(0) \, J^y_{s,\mu}(t) \big\rangle + \big\langle J^z_{s,\nu}(0) \, J^z_{s,\mu}(t) \big\rangle \Big), 
\end{eqnarray}
instead of Eq. (\ref{eq:conductivity}). The spin-current conductivity so obtained is shown in Fig. \ref{fig:spin_comp} (c) as a function of temperature. In the Heisenberg case, neither a magnetic transition nor a topological one does not occur, so that the characteristic temperature scale is absent except the exchange interaction $|J|$. In Fig. \ref{fig:spin_comp} (c), with decreasing temperature, the longitudinal spin-current conductivity $\sigma^s_{xx}$ increases monotonically and a steep increase sets in around $T/|J|\sim 0.8$. As the system size dependence of $\sigma^s_{xx}$ becomes considerably larger at lower temperatures, we will extrapolate the low-temperature $\sigma^s_{xx}(T)$ curve in the thermodynamic limit.

\begin{figure}[t]
\includegraphics[scale=1.0]{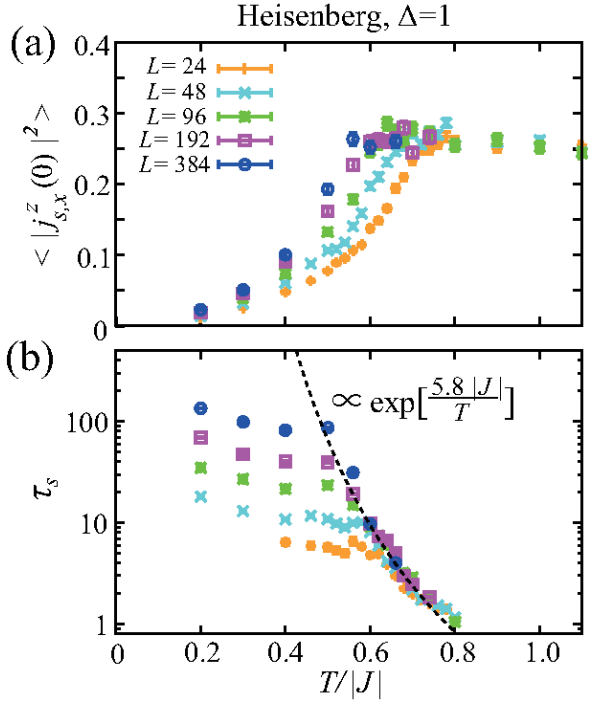}
\caption {The temperature dependences of the equal-time spin-current correlation $\langle | j^z_{s,x}(0)|^2 \rangle$ (a), and the relaxation time of the spin current $\tau_s$ (b), in the Heisenberg case of $\Delta = 1$, where these quantities are measured in the same units as those in Fig. \ref{fig:spin_tau_Ising}. A dashed curve in (b) represents an exponential function obtained by fitting the size-independent data at $0.54 \leq T/|J| \leq 0.74$. \label{fig:spin_tau_Heisenberg}}
\end{figure}

Figure \ref{fig:spin_Linf_Heisenberg} (a) shows the system-size dependence of $\sigma^s_{xx}$ at various temperatures. 
At lower temperatures, a larger system size is necessary to obtain the thermodynamic-limit value of $\sigma^s_{xx}$, suggesting that in contrast to the thermal transport which is a spatially local phenomenon, the spin transport captures the long-length-scale magnetic properties. The extrapolated thermodynamic-limit values of $\sigma^s_{xx}$ and the corresponding original finite-size data in Fig. \ref{fig:spin_comp} (c) are plotted in Fig. \ref{fig:spin_Linf_Heisenberg} (b) on the semi-logarithmic scale as a function of the inverse temperature $1/T$. As the $L \rightarrow \infty$ data at $T/|J| \lesssim 0.8$ are on a straight line, we fit them by the exponential function of $A_H \exp\big[ B_H |J|/T \big]$ with $A_H$ and $B_H$ being fitting parameters. 
Note that in the Heisenberg model in two dimensions, $\xi_s$ increases in the exponential form of $\xi_s/a \sim \exp\big[ b_H |J|/T \big]$ with $b_H \simeq 2 \pi$ \cite{Heisenberg_Polyakov_75}.
The resultant fitting function with the obtained values of $A_H=0.0017 \pm 0.0003$ and $B_H=5.1 \pm 0.1$ is represented by a dashed curve in Figs. \ref{fig:spin_Linf_Heisenberg} (b) and \ref{fig:spin_comp} (c). Since the obtained value of $B_H \simeq 5.1$ is comparable to $b_H \simeq 2\pi$, it turns out that $\sigma^s_{xx} \propto \xi_s$, which is in good agreement with the analytical result in Eq. (\ref{eq:conductivity_classical_spin}). To get insight into the origin of the rapid increase of $\sigma^s_{xx}$, we examine the temperature dependences of $\langle | j^z_{s,x}(0)|^2 \rangle$ and $\tau_s$ like in the anisotropic cases of $\Delta \neq 1$.   

The temperature dependences of $\langle | j^z_{s,x}(0)|^2 \rangle$ and $\tau_s$ are shown in Figs. \ref{fig:spin_tau_Heisenberg}, where $\tau_s$ is extracted from $\langle  j^z_{s,x}(0) \, j^z_{s,x}(t) \rangle$ in the same way as before. $\langle | j^z_{s,x}(0)|^2 \rangle$ is size dependent even at the lowest temperature, but its temperature dependence is relatively weak. 
In LSWT, as shown in Eq. (\ref{eq:Js_static}), $\xi_s$ becomes relevant at lower temperatures, so that $L \gg \xi_s$ should be satisfied to evaluate the thermodynamic limit of $\langle | j^z_{s,x}(0)|^2 \rangle$. As is suggested from the size dependent data, however, the maximum size of $L=384$ seems to be still small and the expected temperature dependence of $const \, T^2 + T$ cannot be seen. Compared with $\langle | j^z_{s,x}(0)|^2 \rangle$, the temperature dependence of $\tau_s$ is much more remarkable.  
As one can see from Fig. \ref{fig:spin_tau_Heisenberg} (b), $\tau_s$ gets longer rapidly toward $T=0$, showing the considerably large system-size dependence. We fit the almost size-independent data at $0.54 \leq T/|J| \leq 0.74$ with the exponential form of $\tilde{A}_H \exp\big[ \tilde{B}_H |J|/T \big]$. The resultant curve with the obtained values of the fitting parameters $\tilde{A}_H=0.0006 \pm 0.0002$ and $\tilde{B}_H=5.8 \pm 0.2$ is represented by a dashed curve in Fig. \ref{fig:spin_tau_Heisenberg} (b). The obtained value of $\tilde{B}_H \simeq 5.8$ is close to $b_H\simeq 2\pi$ and $B_H \simeq 5.1$. 
As $\sigma^s_{xx}$ is estimated roughly as $\sigma^s_{xx} \simeq T^{-1} \, \langle |j^z_{s,x}(0)|^2\rangle \, \tau_s$, the origin of the steep increase of $\sigma^s_{xx}$ toward $T=0$ is the enhanced relaxation-time $\tau_s$ which seems to have a direct association with the rapid growth of the spin correlation length $\xi_s /a \sim \exp\big[ b_H |J|/T \big]$.

\section{Summary and discussion}
We have theoretically investigated transport properties of the classical antiferromagnetic XXZ model on the square lattice in which the anisotropy of the exchange interaction $\Delta \equiv J_z/J_x$ plays a role to control the universality class of the system. In Ising-type ($\Delta >1$), $XY$-type ($\Delta <1$), and Heisenberg-type ($\Delta=1$) magnets, spins in the low-temperature phase are, respectively, long-range-ordered via a magnetic phase transition, quasi-long-range-ordered via the KT topological transition, and disordered. Based on the linear response theory, we have calculated the thermal conductivity $\kappa_{\mu\nu}$ and the spin-current conductivity $\sigma^s_{\mu\nu}$ by means of the hybrid Monte-Carlo and spin-dynamics simulations. It is found that $\sigma^s_{\mu\nu}$ reflects the effect of the anisotropy, i.e., the difference in the ordering properties, while $\kappa_{\mu\nu}$ does not with its longitudinal component $\kappa_{xx}$ (=$\kappa_{yy}$) increasing toward $T=0$ as a power function of temperature independently of $\Delta$. For the $XY$-type anisotropy, the longitudinal spin-current conductivity $\sigma^s_{xx}$ ($=\sigma^s_{yy}$) exhibits a divergence at the Kosterlitz-Thouless (KT) transition temperature $T_{KT}$ obeying the exponential form, $\sigma^s_{xx} \propto \exp\big[ B/\sqrt{T/T_{KT}-1 }\, \big]$ with $B={\cal O}(1)$, while for the Ising-type anisotropy, the temperature dependence of $\sigma^s_{xx}$ is almost monotonic without showing a clear anomaly at the magnetic transition temperature $T_{N}$. In the Heisenberg-type isotropic case, $\sigma^s_{xx}$ exhibits a monotonic exponential increase toward $T=0$. By analyzing the time correlation of the spin current at various temperatures, we find that the divergent enhancement of $\sigma^s_{xx}$ at $T_{KT}$ is due to the exponential rapid growth of the spin-current-relaxation time toward $T_{KT}$. Such a long spin-current-relaxation time can be interpreted as a manifestation of the topological nature of a vortex whose lifetime is expected to get longer toward $T_{KT}$ since the pair-annihilation of vortices should occur more sporadically with the increase of the inter-free-vortex distance toward $T_{KT}$. This suggests that the topological object of the vortex excitation should be crucial for the spin transport. 

Now, we will address possible experimental platforms to investigate the pronounced enhancement of the longitudinal spin-current conductivity $\sigma^s_{\mu \mu}$ associated with the KT transition. As the divergent peak in the $\sigma^s_{\mu\mu}(T)$ curve toward $T_{KT}$ can commonly be seen in both ferromagnets and antiferromagnets only if an $XY$-type anisotropy exists, good candidate systems are quasi-two-dimensional magnets having the signature of the KT transition such as the $S=1/2$ square-lattice ferromagnet K$_2$CuF$_4$ \cite{KCuF_Hirakawa_jpsj_81, KCuF_Hirakawa_jpsj_82, KCuF_Hirakawa_JAP_82, KCuF_Sachs_prb_13}, the $S=1$ honeycomb-lattice antiferromagnets BaNi$_2$X$_2$O$_8$ (X=As, P, V) \cite{BaNiXO_Regnault_JMMM_80, BaNiXO_Regnault_PhysicaB_86, BaNiPO_Regnault_PhysicaB_89, BaNiPO_Gaveau_JAP_91, BaNiVO_Heinrich_prl_03, BaNiVO_Waibel_prb_15, BaNiVO_Klyushina_prb_17}, the $S=5/2$ honeycomb-lattice antiferromagnet MnPS$_3$ \cite{MnPS_Wildes_JPCM_98, MnPS_Ronnow_PhysicaB_00, MnPS_Wildes_prb_06}, and the stage-2 NiCl$_2$ \cite{NiCl_Karimov_JETP_74, NiCl_Karimov_JETP_75, NiCl_Ikeda_JPC_81} and CoCl$_2$ \cite{CoCl_Matsuura_JMMM_83, CoCl_Ikeda_jpsj_85, CoCl_Wiesler_Zphys_94} graphite intercalation which are respectively $S=1$ and $S=1/2$ triangular-lattice ferromagnets. In these compounds, a three-dimensional inter-layer coupling is extremely small, so that at first sight, the system may be regarded as a two-dimensional $XY$-type magnet. In reality, however, on approaching $T_{KT}$ at which the spin correlation length $\xi_s$ diverges, the effective coupling between neighboring layers grows rapidly as the area of the correlated region $\xi_s^2$ rapidly increases, eventually leading to a three-dimensional long-range-order as long as such a perturbative coupling is nonzero. Indeed, all the above compounds undergo a phase transition into a long-range-ordered state before reaching $T_{KT}$. Nevertheless, they have a two-dimensional $XY$-like crossover regime just above the magnetic transition, in which the critical phenomena peculiar to the KT transition have been observed. Thus, measurements of the spin-current conductivity in this crossover regime could, in principle, detect the pronounced enhancement of the longitudinal spin-current conductivity toward the virtually existing $T_{KT}$. 

In the $XY$ magnets, the true divergence associated with the topological transition cannot be detected because the three-dimensional long-range-order inevitably appears before reaching $T_{KT}$. In Heisenberg magnets, however, such a divergence might be detectable if there exists a magnetic frustration leading to a non-collinear spin-ordering. In such frustrated Heisenberg magnets, a topological defect is the so-called $Z_2$ vortex and the KT-type $Z_2$-vortex transition is expected to occur at $T_v$ \cite{Z2_Kawamura_84,Z2_Kawamura_10, Z2_Kawamura_11}. In contrast to the KT transition, although the inter-free-vortex distance diverges at $T_{v}$, $\xi_s$ remains finite at any finite temperature. Thus, a divergent enhancement associated with the $Z_2$-vortex transition, if it occurs, is not necessarily masked by a three-dimensional long-range-order in real materials. This may be an interesting issue, but we will leave further detailed analysis for our future work.

As demonstrated in the present paper, the thermal transport is insensitive to the difference in the ordering properties. In extracting the magnetic contribution from the total longitudinal thermal conductivity, great care has to be taken because it contains phonon contribution as well in the temperature range typical for magnetic transitions. In contrast, the spin-current conductivity should be of purely magnetic origin unless a magnon-phonon coupling is strong enough, suggesting that the spin-current measurements may be a promising probe to detect nontrivial magnetic excitations such as vortices.

\begin{acknowledgments}
The authors thank K. Uematsu, S. Furuya, and Y. Niimi for useful discussions. We are thankful to ISSP, the University of Tokyo for providing us with CPU time. This work is supported by JSPS KAKENHI Grant Numbers JP16K17748, JP17H06137.
\end{acknowledgments}

\appendix
\section{Ordering properties of the classical antiferromagnetic XXZ model on the square lattice}
\begin{figure*}[t]
\includegraphics[scale=0.78]{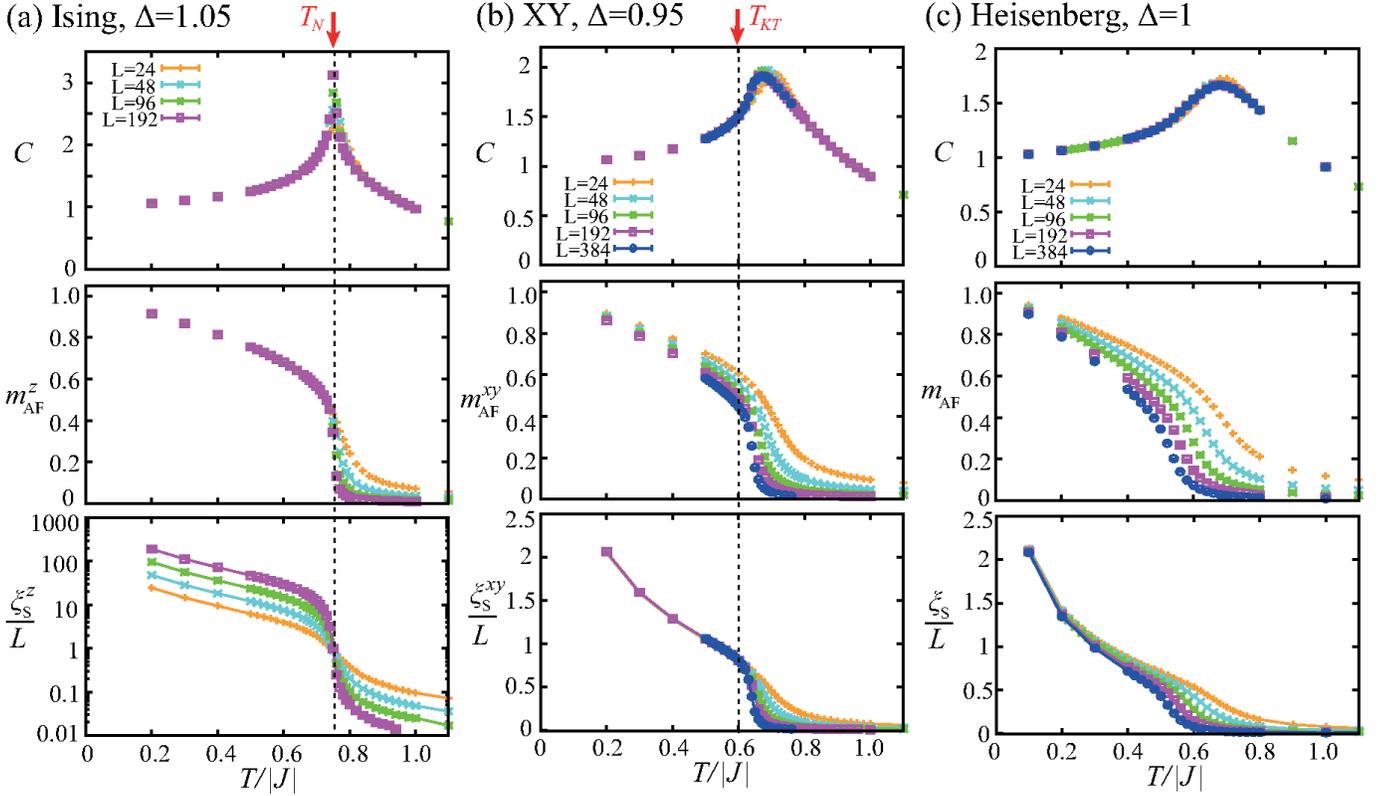}
\caption{The temperature dependences of the specific heat (upper), the antiferromagnetic order parameter (middle), and the ratio of the spin correlation length to the system size (bottom) in the (a) Ising-type ($\Delta=1.05$), (b) $XY$-type ($\Delta=0.95$), and (c) Heisenberg-type ($\Delta=1$) spin systems. In (a) and (b), red arrows indicate the magnetic and KT transition temperatures, $T_N/|J|\simeq 0.75$ and $T_{KT}/|J| \simeq 0.6$, respectively. \label{fig:basic_comp}}
\end{figure*}

The ordering properties of the classical antiferromagnetic XXZ model (\ref{eq:Hamiltonian}) on the square lattice can be investigated by MC simulations. Figure \ref{fig:basic_comp} shows the temperature dependences of the specific heat $C$ (upper), the order parameter for the two-sublattice antiferromagnetic order (middle), the ratio of the spin-correlation length to the linear system size $L$ (bottom) for $\Delta=1.05$, 0.95, and 1. Here, in the Ising-type ($\Delta=1.05$), $XY$-type ($\Delta=0.95$), and Heisenberg-type ($\Delta=1$) spin systems, the order parameters and the associated spin-correlation lengths are respectively given by $m_{\rm AF}^z$ and $\xi_s^z$, $m_{\rm AF}^{xy}$ and $\xi_s^{xy}$, and $m_{\rm AF}$ and $\xi_s$ which are defined by 
\begin{eqnarray}
m_{\rm AF}^z &=& \sqrt{ G^z({\bf Q}) }, \nonumber\\
m_{\rm AF}^{xy} &=& \sqrt{ G^x({\bf Q}) + G^y({\bf Q}) }, \nonumber\\
m_{\rm AF} &=& \sqrt{ G^x({\bf Q}) + G^y({\bf Q}) + G^z({\bf Q}) } \nonumber\\
\xi_s^z &=& \frac{1}{2\sin(\pi/L)} \sqrt{\frac{G^z({\bf Q})}{G^z({\bf Q}+{\bf k}_{\rm min})}-1 }, \nonumber\\
\xi_s^{xy} &=& \frac{1}{2\sin(\pi/L)} \sqrt{\frac{\sum_{\alpha=x,y}G^\alpha({\bf Q})}{\sum_{\alpha=x,y}G^\alpha({\bf Q}+{\bf k}_{\rm min})}-1 }, \nonumber\\
\xi_s &=& \frac{1}{2\sin(\pi/L)} \sqrt{\frac{\sum_{\alpha=x,y,z}G^\alpha({\bf Q})}{\sum_{\alpha=x,y,z}G^\alpha({\bf Q}+{\bf k}_{\rm min})}-1 }, \nonumber\\
G^\alpha({\bf q}) &=& \big\langle \big|\frac{1}{N_{\rm spin}}\sum_i S^\alpha_i \, e^{i \, {\bf q}\cdot {\bf r}_i}\big|^2 \big\rangle,\nonumber\\
{\bf Q} &=& (\pi, \pi), \qquad {\bf k}_{\rm min} = (2\pi/L,0).
\end{eqnarray}
In our MC simulations, we perform $3\times 10^5$ MC sweeps and the first $10^5$ sweeps are discarded for thermalization, where one MC sweep consists of the 1 heat-bath sweep and successive 10-30 over-relaxation sweeps. Observations are done in every MC sweep, and the statistical average is taken over 10 independent runs starting from different initial spin configurations.  

As one can see from Fig. \ref{fig:basic_comp} (a), in the Ising case of $\Delta=1.05$, the specific heat $C$ exhibits a sharp peak associated with the antiferromagnetic transition at $T_N/|J|\simeq 0.75$. Correspondingly, $m_{\rm AF}^z$ starts growing up at $T_N$ and $\xi_s^z/L$ for different system sizes cross one another at $T_N$, which is usually the case for ordinary continuous magnetic phase transitions.

In the $XY$ case of $\Delta=0.95$, the KT transition temperature is estimated to be $T_{KT}/|J| \simeq 0.6$ in Refs. \cite{KT_XXZ_Cuccoli_95, KT_XXZ_Lee_05, KT_XXZ_Pires_96}. Actually, as one can see from Fig. \ref{fig:basic_comp} (b), $\xi_s^{xy}/L$ for different system sizes merge one another below $T_{KT}$, whereas the specific heat only shows a broad peak slightly above $T_{KT}$ and $m_{\rm AF}^{xy}$ is suppressed with increasing $L$ because of the absence of the true magnetic long-range order. 

In the Heisenberg case of $\Delta=1$, the specific heat shows only a broad peak near $T/|J|\simeq 0.7$ and the spin-correlation length $\xi_s$ is finite at any finite temperature as is suggested from the fact that in Fig. \ref{fig:basic_comp} (c) $\xi_s/L$ continues to be suppressed with increasing the system size $L$ at all the temperatures.

\end{document}